\documentclass[%
 reprint,
superscriptaddress,
 amsmath,amssymb,
aps,
]{revtex4-2}

\usepackage{graphicx}
\usepackage{dcolumn}
\usepackage{bm}

\usepackage{pifont} 
\usepackage{xcolor} 
\usepackage{makecell}
\usepackage[colorlinks=true,citecolor=blue,linkcolor=blue,pdfhighlight =/O]{hyperref}

\newcommand{\eref}[1]{Eq.~(\ref{#1})}

\newcommand{\sfref}[1]{Figure~\ref{#1}}

\newcommand{\fref}[1]{Fig.~\ref{#1}}

\hypersetup{urlcolor=blue}

\begin{document}

\preprint{APS/123-QED}

\title{Photothermal responsivity of van der Waals material-based nanomechanical resonators}

\author{Myrron Albert C. Aguila}
\email{maguila@gate.sinica.edu.tw}
\affiliation{Department of Engineering and System Science, National Tsing Hua University, Hsinchu, 30013, Taiwan}
\affiliation{Nano Science and Technology Program, Taiwan International Graduate Program, National Tsing Hua University and Academia Sinica}
\affiliation{Institute of Physics, Academia Sinica, Nankang, Taipei, 11529, Taiwan}

\author{Joshoua C. Esmenda}%
\email{jesmenda@gate.sinica.edu.tw}
\affiliation{Department of Engineering and System Science, National Tsing Hua University, Hsinchu, 30013, Taiwan}
\affiliation{Nano Science and Technology Program, Taiwan International Graduate Program, National Tsing Hua University and Academia Sinica}
\affiliation{Institute of Physics, Academia Sinica, Nankang, Taipei, 11529, Taiwan}

\author{Jyh-Yang Wang}
\affiliation{Institute of Physics, Academia Sinica, Nankang, Taipei, 11529, Taiwan}

\author{Teik-Hui Lee}
\affiliation{Institute of Physics, Academia Sinica, Nankang, Taipei, 11529, Taiwan}

\author{Yen-Chun Chen}
\affiliation{Institute of Physics, Academia Sinica, Nankang, Taipei, 11529, Taiwan}

\author{Chi-Yuan Yang}
\affiliation{Institute of Physics, Academia Sinica, Nankang, Taipei, 11529, Taiwan}

\author{Kung-Hsuan Lin}
\affiliation{Institute of Physics, Academia Sinica, Nankang, Taipei, 11529, Taiwan}

\author{Kuei-Shu Chang-Liao}
\affiliation{Department of Engineering and System Science, National Tsing Hua University, Hsinchu, 30013, Taiwan}

\author{Sergey Kafanov}
\affiliation{Department of Physics, Lancaster University, Lancaster, LA1 4YB, United Kingdom}

\author{Yuri A. Pashkin}
\affiliation{Department of Physics, Lancaster University, Lancaster, LA1 4YB, United Kingdom}

\author{Chii-Dong Chen}
\email{chiidong@phys.sinica.edu.tw}
\affiliation{Institute of Physics, Academia Sinica, Nankang, Taipei, 11529, Taiwan}


\date{\today}

\begin{abstract}

Nanomechanical resonators made from van der Waals materials (vdW NMRs) provide a new tool for sensing absorbed laser power. The photothermal response of vdW NMRs, quantified from the resonant frequency shifts induced by optical absorption, is enhanced when incorporated in a Fabry-Perot (FP) interferometer. Along with the enhancement comes the dependence of the photothermal response on NMR displacement, which lacks investigation. Here, we address the knowledge gap by studying electromotively driven niobium diselenide drumheads fabricated on highly reflective substrates. We use a FP-mediated absorptive heating model to explain the measured variations of the photothermal response. The model predicts a higher magnitude and tuning range of photothermal responses on few-layer and monolayer NbSe$_{2}$ drumheads, which outperform other clamped vdW drum-type NMRs at a laser wavelength of $532\,$nm. Further analysis of the model shows that both the magnitude and tuning range of NbSe$_{2}$ drumheads scale with thickness, establishing a displacement-based framework for building bolometers using FP-mediated vdW NMRs.
\end{abstract}


\maketitle


\section{Introduction}

Nanomechanical resonators (NMRs) embedded in an optical cavity are valuable platforms for studying weak forces due to the enhanced coupling between light and motion\cite{Aspelmeyer2014CavityOpt}. Enhanced coupling improves the capability of nanomechanical resonators to demonstrate nonlinear dynamics\cite{Davidovikj2017Nonlinear} and sense heat transport in suspended nanostructures\cite{Morell2019MoSe2}. Resonators interacting with optical elements enjoy additional degree of spatiality\cite{Waitz2012Spa,Wang2014Spectromicro,Davidovikj2016Visual,Kim2018,esmenda:2020}, extremely large optomechanical coupling at ambient temperature\cite{Purdy2017QuantCorr,Delic2020RTCooling}, and reduced mode volume due to breakthrough technologies in focusing laser beams via free space optics\cite{bagci2014optical,andrews2014bidirectional,Higginbotham2018EO}, fiber optics\cite{Azak2007Fiber,Metzger2008pth,FJ2012fpopt} and near-field interactions with multiplexed on-chip optical waveguides and tapered fibre\cite{Basarir2012MotionTrans,Vainsencher2016Bidirec}.

Heating induced by the probe laser remains a concern in the optical readout of NMRs\cite{Midolo2018NOEMS,chen2018vibration}. Photothermal effects\cite{Metzger2008pth,Barton2012pth,Morell2019MoSe2,Primo2021pthmodelling} emerge once the laser illuminates a region of the resonator and raises the temperature of the mechanical mode\cite{chen2018vibration}. While the process hinders ground-state cooling of the mechanical mode\cite{Kippenberg2008CavFeed}, it enables sensing of incident laser power with the aid of on-chip Fabry-Perot (FP) cavities. NMRs fabricated with suspended van der Waals (vdW) materials\cite{Barton2011Q,Lemme2020NEMS,Steeneken2021Dynamics} show promising photothermal sensitivities\cite{Blaikie2019GrBol} due to reduced mass, and layer-dependent mechanical, optical, and thermal properties. Few-layer and multilayer niobium diselenide (NbSe$_{2}$) is a candidate vdW material for low-noise, highly responsive photodetectors at ultraviolet\cite{Hu2020UVphoto} and infrared wavelengths\cite{Orchin2019,Jin2021CryoNbSe2}.

There are several approaches to quantifying the power absorbed by vdW NMRs. One can measure the mechanical mode temperature by resolving thermomechanical motion\cite{Davidovikj2016Visual,Papas2021omn} while varying the incident laser power. Such detection requires low mass and high quality factors of NMRs, which can be difficult to ensure for vdW materials\cite{Barton2011Q,Lemme2020NEMS,Steeneken2021Dynamics}. A less stringent yet popular method involves monitoring of the resonant frequency shift of an electromotively driven vdW NMRs\cite{Morell2019MoSe2,Blaikie2019GrBol} and following its dependence on the incident laser power. In both approaches, light absorption depends not only on the FP structure and wavelength, but also on the resonator displacement from the initial equilibrium position. Understanding the displacement-dependent absorptive response may provide insights into managing FP-mediated heat flow in NMRs. While experiments on NMRs fabricated from various types of vdW materials have been published, photothermal sensing with NbSe$_{2}$ NMRs has not yet been reported. NbSe$_{2}$ has low thermal conductivity\cite{Ferreiro2019}, is flexible\cite{sengupta2010electromechanical} and has large fracture strain\cite{Sun2021NbSe2}. This combination of properties enables the realization of low power, photothermal-strain-sensitive detectors.

In this paper, we explore the influence of the resonator displacement on the sensitivity of the FP-based vdW NMR to incident light. We investigate the photothermal response of electromotively driven NbSe$_{2}$ drumheads. We propose an FP-mediated absorptive heating model that accounts for resonator displacement to explain the observed variations in the measured photothermal response. The model consequently reveals a large tuning range of photothermal responsivity as the drumhead moves towards the bottom electrodes. We extend the model to drumheads of varying thicknesses and materials to gain insight into the geometric and material impact of FP-mediated heating on NMRs.

\begin{figure}[htb]
\includegraphics{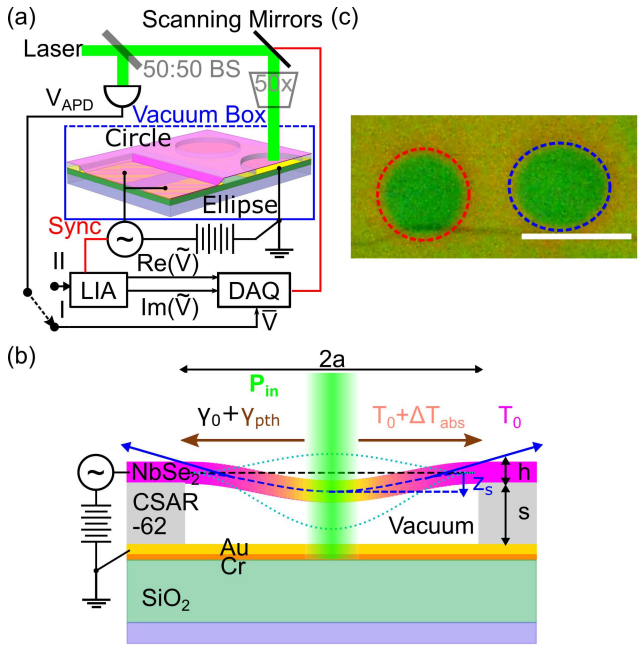}
\caption{\label{fig:Intro} Measurement setup and device. \textbf{(a)} Optical interferometric setup used to track the mechanical frequency of multilayered NbSe$_{2}$ flake mechanical resonators. \textbf{(b)} Schematic diagram of laser-induced photothermal heating of an electromotively driven drumhead resonator. Cross-section showing the net tension exerted by the drumhead resonators under photothermal heating. \textbf{(c)} Optical micrograph of the circular (red dashed circle) and elliptical (blue dashed ellipse) drumhead resonators under study. The white scale bar corresponds to a length of 10 $\mu$m.}
\end{figure}

\section{Materials and Methods}

The mechanical drums are set into motion by a combination of static and oscillating voltage. This motion is detected through the optical interferometric detection scheme at a laser wavelength $\lambda=532\,$nm, as shown in \fref{fig:Intro}(a) and (b). When a laser beam illuminates a region of the drumhead, the drum surface reflects a fraction of the incident power $P_{in}$ for detection and absorbs some of the power accumulated in the FP cavity. The detector receives the modulated reflected beam needed to resolve the driven frequency response of the drumheads. The absorbed power can be described as $P_{abs}=P_{in} \chi A_{FP}$, where $A_{FP}$ is the total absorbance of the FP cavity and $\chi$ is the power fraction absorbed by the drumhead in the FP stack. The absorbance of the FP cavity depends on the difference between the spacer height $s$ and the resonator static displacement $z_{s}$. $P_{abs}$ heats the illuminated spot, and produces a temperature gradient due to the radial heat transfer to the drum clamps, as shown in \fref{fig:Intro}(b). Consequently, the elevated temperature changes the mechanical tension exerted by the clamps by photothermal tension $\gamma_{pth}$. As illustrated in \fref{fig:Intro}(b), $\gamma_{pth}$ is added to the existing initial mechanical tension $\gamma_{0}$ generated by the drum displacement. 

\sfref{fig:Intro}(c) shows the optical micrograph of devices A (circular) and B (elliptical). The drumheads are made from exfoliated NbSe$_{2}$ flakes, which are suspended on a lithographically patterned Au-Cr-SiO$_{2}$-Si substrate covered with electron-beam patterned resist CSAR-62. Device A has a radius $a_{A}=b_{A}=3.5\,\mu$m and device B has a major radius $a_{B}=4.0\,\mu$m and a minor radius $b_{B}=3.5\,\mu$m. Details of the fabrication and optical detection are explained elsewhere\cite{esmenda:2020}. Using the multilayer interface approach (MIA)\cite{Aguila2022FPCal}, the spacer height is determined to be $s=296\,$nm and the drumhead thickness $h=55\,$nm. Observation of the photothermal effect on the mechanical drums requires changing the laser power and monitoring the mechanical frequency shift. 

\begin{figure}[htb]
\includegraphics{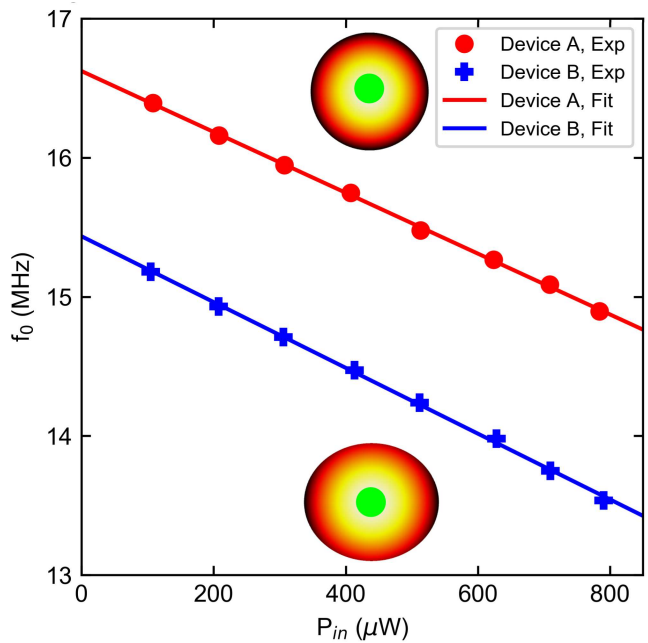}
\caption{\label{fig:Results} Dependence of the fundamental resonant frequencies of circular (device A) and elliptical (device B) resonators on the incident laser power. Insets show the position of the laser spot (green circle in the center) at which the driven responses of devices A and B are measured. The devices are electromotively driven at $V_{DC}=4\,$V and $V_{AC}=0.125\,$V$_{pk}$. Data points are measured resonant frequencies from the driven responses and solid lines are linear fits.}
\end{figure}

\section{Results and Discussion}

We consider the case of a tensioned circular drumhead where both flexural rigidity $D_{p}$ and $\gamma_{0}$ have comparable effect on the resonant frequency of the drumhead. The resonant frequency of both circular and elliptical drumheads, with an effective radius $a_{eff}=\sqrt{ab}$ and thickness $h$ can be written as\cite{Lee2013}
\begin{equation}
    {
     f_{0}(P_{in})= \frac{\lambda_{01}}{2 \pi}\sqrt{\frac{ D_{p} }{\rho h a_{eff}^{4}} \left[ \lambda_{01}^2 + \frac{{{(\gamma_{0}+ \gamma_{pth})}{a_{eff}^2}}}{{D_{p}}} \right]},  
    }    
    \label{eqn:f_tens_plate}
\end{equation}
where $\lambda_{01}$ is a modal parameter that is determined numerically. At $P_{in}=0$, the drumheads oscillate at their natural resonant frequency (without heating) $f_{0}=f_{0}(P_{in}=0)$ corresponding to the y-intercept of both plots for circular and elliptical drumheads in \fref{fig:Results}. Given $f_{0}$ values of 16.621 and 15.432 MHz for device A and B, respectively, and using \eref{eqn:f_tens_plate}, we determine the initial tensions of $\gamma_{0,A}=5.01\,$N$\,$m$^{-1}$ for device A and $\gamma_{0,B}=5.15\,$N$\,$m$^{-1}$ for device B, for the given applied DC voltage. As $P_{in}$ ramps up, downward resonant frequency shift is observed as shown in \fref{fig:Results}. Since the frequency shift is linear for small values of $P_{in}$, \eref{eqn:f_tens_plate} can be given by its first-order Taylor polynomial 
\begin{equation}
	{
	f_{0}(P_{in}) \approx f_{0} + \frac{1}{2} \left[ \left( \frac{\lambda_{01}}{2 \pi} \right)^{2} \frac{1}{ \rho h a_{eff}^{2} } \right] \frac{\gamma_{pth}(P_{in})}{f_{0}}
	}
	\label{eqn:f_taylor}
\end{equation}
and the shift can be written as
\begin{equation}
	{
	\Delta f_{0}(P_{in}) = \frac{1}{2} \left( \frac{\lambda_{01}}{2 \pi} \right)^{2} \frac{\gamma_{pth}(P_{in})}{\rho h a_{eff}^{2} f_{0}}.
	}
	\label{eqn:del_fm_simp}
\end{equation}
The compressive tension is given by\cite{Blaikie2019GrBol,Siskins2020}
\begin{equation}
	{
	\gamma_{pth}(P_{in}) = - \frac{E_{3D} h}{1- \nu} \alpha_{L} \Delta T_{abs}(P_{in}),
	}
	\label{eqn:radial_tension_heat}
\end{equation}
where $E_{3D}$ is the Young's elastic modulus, $\nu$ is the Poisson's ratio, $\alpha_{L}$ is the thermal expansion coefficient of NbSe$_{2}$ at the bath temperature $T_{0}$, and $\Delta T_{abs}$ is the temperature difference between $T_{0}$ and the average temperature of the drumhead $T_{0} + \Delta T_{abs}$. For linear changes in $T_{0}+\Delta T_{abs}$, the average temperature difference is expressed as\cite{Ye2018Therm}
\begin{equation}
	{
	\Delta T_{abs}(P_{in}) = \frac{P_{abs}}{4 \pi h \kappa} \eta,
	}
	\label{eqn:Delta_T_simp}
\end{equation}
where $\kappa$ is the in-plane thermal conductivity of NbSe$_{2}$, and $\eta$ is the average spot size factor. We estimate both $\chi$ and $A_{FP}$ through MIA\cite{BaumeisterOCT,Islam2018Ani,ByrnesMOC} whereas $\eta$ is evaluated by assuming absorptive spot heating in the center of the drumheads\cite{Kurek2017Heat}.

Given these inputs, we define the photothermal responsivity $\Psi$ as the frequency shift induced by the absorbed power. By solving \eref{eqn:del_fm_simp} using \eref{eqn:radial_tension_heat} and \eref{eqn:Delta_T_simp}, and defining $\Psi$ as $\Psi=\Delta f_{0} / \Delta P_{in}$, the photothermal responsivity is expressed as
\begin{equation}
{
	\Psi(z_{s}) = - \frac{1}{8} \left( \frac{\lambda_{01}}{2 \pi} \right)^{2} \frac{E_{3D} \alpha_{L}}{(1-\nu) m f_{0}} \frac{\chi A_{FP} \eta}{\kappa}, 
} \label{eqn:pth_static}
\end{equation}
where $m= \rho \pi a_{eff}^2 h$ is the total mass of the drumhead. This quantity can be extracted experimentally from the slope of the linear fits acquired from \fref{fig:Results}.

\begin{figure}[htb]
\includegraphics{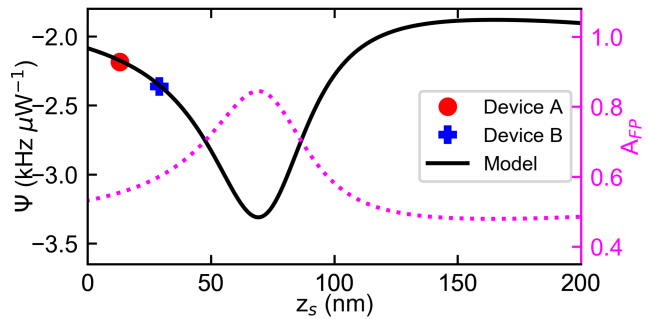}
\caption{\label{fig:Photothermal Model} Photothermal responsivity dependence of the multilayer NbSe$_{2}$ drumhead resonators on the static displacement. Data points are the slope extracted from the slope in \fref{fig:Results}. Solid line refers to the photothermal responsivity model. For comparison, the dependence of the simulated absorbance of the Fabry-Perot cavity on static displacement, shown by magenta dotted line with the magnitude referenced on the right Y-axis spine, is also shown.}
\end{figure}

\begin{figure*}
\includegraphics{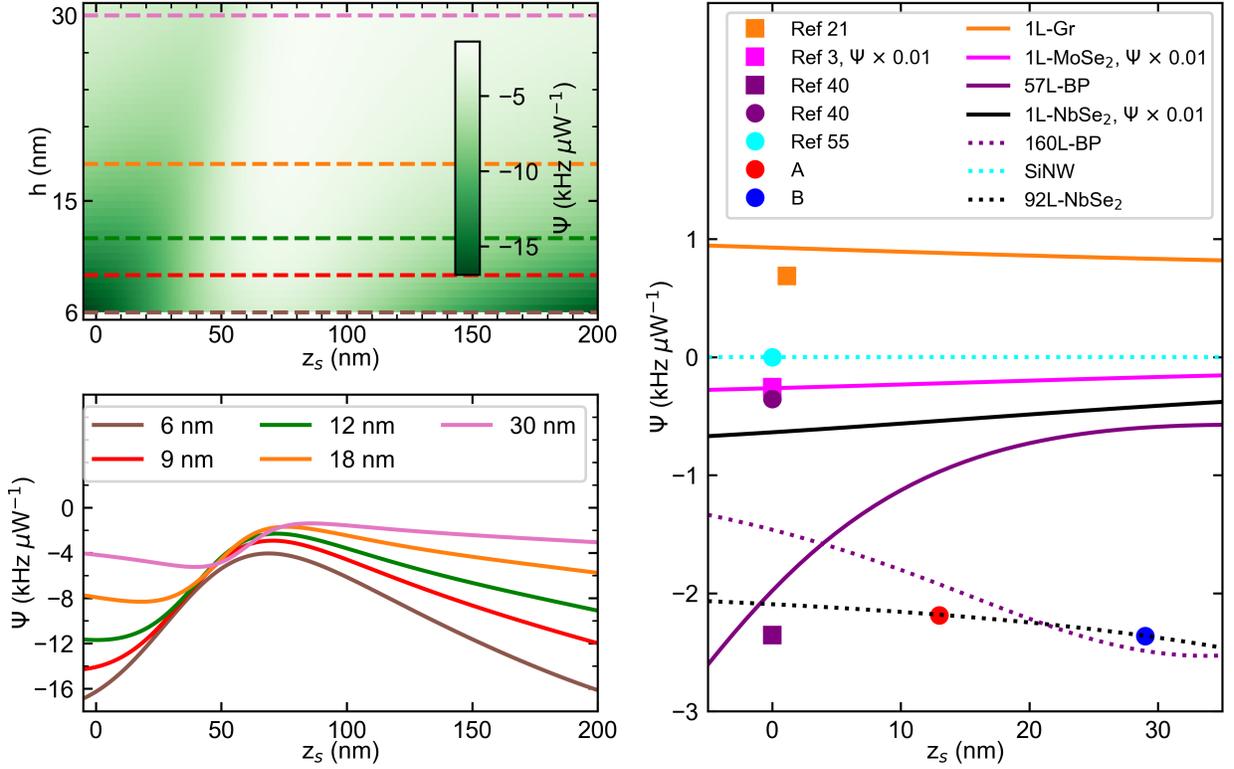}
\caption{\label{fig:Comparison}Dependence of the displacement-dependent photothermal responsivity profile $\Psi (z_{s})$ of drum NMRs on the thickness and resonator material. \textbf{(a)} Intensity plot of the calculated dependence of $\Psi (z_{s})$ of bulk NbSe$_{2}$ drumheads on varying thickness. Colored dashed lines refer to representative $\Psi (z_{s})$ profiles displayed in \textbf{(b)}. For comparison, we use the diameter, Fabry-Perot structure, and mechanical resonant frequency of Device A in simulating the dependences. \textbf{(c)} Scatter plot of the photothermal responsivity values obtained in \fref{fig:Photothermal Model}, along with photothermal responsivities of other NMRs extracted from literature. Square symbols refer to devices demonstrating highly tensioned systems like monolayer (1L) graphene (Gr), monolayer molybdenum diselenide (MoSe$_{2}$), and thin black phosphorus (57L-BP) drumheads. Circular symbols refer to devices dominated by flexural rigidity like thick black phosphorus (160L-BP) drums, silicon nanowire (SiNW) cantilever beams, and NbSe$_{2}$ drums (A, B). Included also are $\Psi (z_{s})$ of both tension-dominated (solid lines) and flexural rigidity-dominated (dotted lines) devices by adopting the Fabry-Perot structure of device A.}
\end{figure*}

\sfref{fig:Photothermal Model} shows the dependence of $\Psi$ extracted from \fref{fig:Results} with the corresponding $z_{s}$ and $\Psi(z_{s})$ generated with \eref{eqn:pth_static} using the resonator specifications of device A. Device A has $\Psi_{A}=-2.19\,$kHz $\mu$W$^{-1}$ for $z_{s}$=13$\,$nm. Device B has $\Psi_{B}=-2.38\,$kHz $\mu$W$^{-1}$ for $z_{s}=29\,$nm. The ratio of $z_s$ between device A and device B in \fref{fig:Photothermal Model} yields 2.23, which is close to the value of 2.16 obtained from the theoretically derived expression $\beta_{ellipse} a_{B}^{4} / \beta_{circle} a_{A}^{4}$, where $\beta$ is the eccentricity factor from reference\cite{Krauthammer}. The results imply that the difference in $z_{s}$ of these two drums lies with geometry\cite{esmenda:2020}. Physically, the compressive strain translates to an out-of-plane radial expansion of the drumhead when aided with $z_{s}$. $z_{s}$ is controlled either through electromotive driving of the drumhead\cite{wong2010characterization,weber2016force,Morell2019MoSe2} or through slack\cite{sazonova2006thesis,wu2011swcnt,esmenda:2020}. Given the nature of the transfer process using PDMS disks\cite{pande2020ultralow,esmenda:2020}, $z_{s}$ likely originates from slack.

According to the model prediction in \fref{fig:Photothermal Model}, \mbox{$\Psi=-2.10$ kHz $\mu$W$^{-1}$} when the resonator is in its equilibrium position ($z_{s}=0$). Displacing the resonator to $z_{s}=69\,$nm would result in \mbox{$\Psi \approx -3.33\,$kHz$\, \mu$W$^{-1}$}, which amounts to increases in the measured responsivities of devices A and B by 42$\%$ and 52$\%$, respectively. Also, we estimate a tuning range of $1.4\,$kHz $\mu$W$^{-1}$. As shown in \fref{fig:Photothermal Model}, the behavior of $A_{FP}$ mimics the observed dependence of $\Psi$ on $z_{s}$, albeit opposite in sign due to heat-induced compression.

We then simulate the effect of varying $h$ on the $\Psi(z_{s})$ profiles of circular NbSe$_{2}$ drumheads using \eref{eqn:pth_static}, as shown in \fref{fig:Comparison}(a,b). The simulated $\Psi(z_{s})$ profiles are restricted to the material properties of bulk, clean devices possessing $a_{eff}$, $f_{0}$, FP structure and $\lambda$ of device A. With these parameters, only $\lambda_{01}$ is varied to decrease at an increasing thickness to maintain $f_{0}$. For $h=6\,$nm, we estimate a tuning range of roughly $12.2\,$kHz $\mu$W$^{-1}$ from $\Psi(z_{s}=0)=-16.30\,$kHz $\mu$W$^{-1}$ to $\Psi(z_{s}=69\,$nm$)=-4.05\,$kHz $\mu$W$^{-1}$. For $h=30\,$nm, we obtain a tuning range of $2.78\,$kHz $\mu$W$^{-1}$ from $\Psi(z_{s}=0)=-4.16\,$kHz $\mu$W$^{-1}$ to $\Psi(z_{s}=86\,$nm$)=1.39\,$kHz $\mu$W$^{-1}$. Hence, thinner bulk NbSe$_{2}$ devices have larger tuning range and magnitude as compared to thicker devices. Furthermore, the simulated $\Psi(z_{s})$ profile of monolayer NbSe$_{2}$, shown in \fref{fig:Comparison}(c) as black solid lines, has larger magnitude and tuning ranges than the extrapolated bulk NbSe$_{2}$ $\Psi(z_{s})$ curve (black dotted line).

Intuitively, the change in $h$ modifies the properties of both the FP cavity and the nanomechanical resonators, and consequently affects $\Psi(z_{s})$. In the FP domain, $h$ modifies both $A_{FP}$ and $\chi$. In a non-transmissible FP cavity with known refractive indices of the resonators $\hat{n}_{res}$\cite{hill2018comprehensive,Weber2010Grnk,Hsu2019Thickness-Dependent,Wang2016AdvancesBP,1997EPalik1}, reflectance and absorbance dominate. For example, the deflection-dependent behavior of the bulk NbSe$_{2}$ $A_{FP}$ transitions from an asymmetric, sinusoidal profile at thin layers to a Fano-peak profile at thicker layers, as shown in Fig. S1 of the Supplementary Material. Also, the deflection dependence of monolayer NbSe$_{2}$ $A_{FP}$ maintains an asymmetric, sinusoidal profile due to its large absorption coefficient, which is different compared to bulk NbSe$_{2}$\cite{hill2018comprehensive,Aguila2022FPCal}. In device A, the values of $z_{s}$ predicted to have maximum absorbance and magnitude of $\Psi$ would have near-zero modulated reflectance, implying no detectable NMR responses by the FP cavity. Hence the estimates for the tuning range of the drumheads are upper bounds. Nevertheless, the values help confine a range of $z_{s}$ that give detectable driven response and photothermal heating. 

In the nanomechanical resonator domain, the effect of the drum head thickness depends on whether the resonator operates in the low tension or high tension regime (see Supplementary Material for the full expressions). In the regime where rigidity dominates, $\Psi \propto h^{-2}$. In the tensioned membrane regime in which the modelled multilayer and monolayer NbSe$_{2}$ drums in \fref{fig:Comparison}(b-c) reside, $\Psi \propto h^{-0.5}$. Engineering a constant value of $\Psi$ for any value of $z_{s}$ requires thicker drum plates. For devices with larger tuning ranges of $\Psi$, thin NbSe$_{2}$ membranes are preferred, though stress-relief structures offer the possibility of thin plate structures\cite{Zhou2020StrainFree}. 

We note that the $\Psi(z_{s})$ curve of monolayer (1L) NbSe$_{2}$ in \fref{fig:Comparison}(c) represents an upper bound by assuming a clean device ($\rho \approx \rho_{NbSe_{2}}$). These conditions result in a greater magnitude and larger tuning range of $\Psi(z_{s})$ of the monolayer NbSe$_{2}$ than $\Psi(z_{s})$ of other devices like 165$\,$nm thick, clamped-free silicon nanowire (SiNW) resonators\cite{Santos2013SiNW} (cyan, solid line), graphene (1L-Gr) (orange, solid line), cryogenically-cooled monolayer diselenide (1L-MoSe$_{2}$) (purple, solid line) and multilayer black phosphorus (BP, magenta solid and dotted lines) drums. The mentioned devices are simulated using the same substrate as Device A. Note that the $\Psi(z_{s})$ profiles for BP resonators were simulated assuming an average, isotropic $E_{3D}$ and $\kappa$ even though these resonators have reported anisotropic properties\cite{Islam2018Ani}. The indicated materials have positive $\alpha_{L}$ which implies that the devices made of these materials experience compression upon the illumination by the laser beam. Only the $\Psi(z_{s})$ curve of graphene experiences tension upon spot heating due to graphene's negative $\alpha_{L}$ at ambient temperature. Furthermore, the SiNW device falls under the optically-thick, bending regime, which illustrates the near-zero magnitude and tunability of $\Psi$ with $z_{s}$. Tables 1, 2, and 3 and Fig. S1 of the Supplementary Material contain more details regarding \fref{fig:Comparison}.

We also see from \fref{fig:Comparison}(c) that the measured $\Psi$ of bulk NbSe$_{2}$ devices is comparable to that of 57L-BP device (magenta square) and significantly better than both bulk SiNW (cyan circle) and 160L-BP (magenta circle) devices. However, $\Psi$'s of these devices are ten times smaller than the $\Psi$ values of a 1L-MoSe$_{2}$ drumhead (pink square), which possesses a significantly smaller mass. We note that the values of $z_{s}$ extracted from literature are for the devices in their equilibrium position except for graphene, which has $1\,$nm deflection to resolve its motion\cite{Barton2012pth}. 

The effect of the substrate and probing wavelength on photothermal heating of the drumheads varies with the resonator material as its optical, thermal, and mechanical properties modify their $\Psi(z_{s})$ profiles. For example, the highly-reflective substrate of device A shows improved $\Psi(z_{s}=0)$ for the 160L-BP, 1L-Gr and 1L-MoSe$_{2}$ device, comparable heating to SiNW device and decreased photothermal heating for the 57L-BP drumhead as compared to their extracted values in literature\cite{Barton2012pth,Morell2019MoSe2,Islam2018Ani,Santos2013SiNW} (colored symbols), as shown in \fref{fig:Comparison}(c). The devices cited in literature make use of a combination of gap heights, probe wavelength, and substrate for near-optimal FP-based detection\cite{wang2016interferometric} and ease in device fabrication; their photothermal responses are secondary. Nevertheless, our model demonstrates the variation of the tuning range of $\Psi(z_{s})$ of thin van der Waals materials using our FP structure. 

The values of $\Psi(z_{s})$ reported in this work, with the exception of clean 1L-NbSe$_{2}$, are fractions of $\Psi$ reported for metamaterial string\cite{Papas2021omn} and graphene trampoline NMR bolometers\cite{Blaikie2019GrBol} at infrared and visible wavelengths, respectively. Apart from replacing NbSe$_{2}$ with thermally insulating vdW materials, a strategy to improve $\Psi(z_{s}=0)$ of NbSe$_{2}$ involves decreasing the thermal conductivity of the drumheads with high-aspect-ratio tethers that resemble a trampoline geometry. The trampoline structure is reported to increase $\Psi(z_{s}=0)$ of graphene, known for its very high thermal conductivity, to $2.4 \times 10^{3}\,$kHz$\, \mu$W$^{-1}$\cite{Blaikie2019GrBol}. The structure is tricky to implement on NbSe$_{2}$ since the focused ion beam etching step introduces defects that alter the flake properties\cite{Li2017FIB,Tomori2019FIB}. On another note, broadening the tuning range would entail engineering a partially-transparent substrate at a probe wavelength to distinguish the resonator positions of maximum absorbance and zero modulated reflectance\cite{chen2018vibration}. 

\section{Conclusions}

In summary, we have designed and characterized NbSe$_{2}$ drumheads with controllable photothermal heating using NMR static displacements in FP cavities. Our simulations show that the magnitude and tuning range of the photothermal response of drumheads increase at decreasing flake thickness. Our analysis shows that a monolayer NbSe$_{2}$ drumhead NMR has promising photothermal responsivities at room temperature. While our work focuses on NbSe$_{2}$ devices, the design framework applies to a family of vdW materials and conventional resonator materials that absorb light. 

\begin{acknowledgments}
We acknowledge the contribution of T.-H. Hsu and W.-H. Chang in fabricating devices and in building the experimental setup, B.-R. Guo for technical assistance, A.F. Rigosi for sharing the measured dielectric constant spectra of bulk and few-layer NbSe$_{2}$ and R. Frisenda for sharing the real and imaginary refractive index of single-layer MoSe$_{2}$. We thank the Taiwan International Graduate Program for the financial support. This project is funded by Academia Sinica Grand Challenge Seed Program (AS-GC-109-08), Ministry of Science and Technology (MOST) of Taiwan (107-2112-M-001-001-MY3), Cost Share Programme (107-2911-I-001-511), the Royal Society International Exchanges Scheme (grant IES$\backslash$R3$\backslash$170029), and iMATE(2391-107-3001). We extend our gratitude for the Academia Sinica Nanocore Facility.
\textit{Attributions.-}M.A.C.A. and J.C.E. contributed equally in this work. C.-D.C. conceived the devices and supervised the project; J.C.E. fabricated the devices. M.A.C.A. and J.-Y.W. modeled the photothermal responsivity dependences. C.-Y.Y. and K.-H.L. designed and built the setup for optical measurements. M.A.C.A., J.C.E., Y.-C.C., and C.-Y.Y. performed the experiment. M.A.C.A., J.C.E., J.-Y.W., Y.-C.C., T.-H.L., K.-S.C.-L., S.K., Y.P. and C.-D.C. analyzed the data, performed simulations and wrote the manuscript; all authors discussed the results and contributed to the manuscript.
\end{acknowledgments}


\begin{thebibliography}{58}%
\makeatletter
\providecommand \@ifxundefined [1]{%
 \@ifx{#1\undefined}
}%
\providecommand \@ifnum [1]{%
 \ifnum #1\expandafter \@firstoftwo
 \else \expandafter \@secondoftwo
 \fi
}%
\providecommand \@ifx [1]{%
 \ifx #1\expandafter \@firstoftwo
 \else \expandafter \@secondoftwo
 \fi
}%
\providecommand \natexlab [1]{#1}%
\providecommand \enquote  [1]{``#1''}%
\providecommand \bibnamefont  [1]{#1}%
\providecommand \bibfnamefont [1]{#1}%
\providecommand \citenamefont [1]{#1}%
\providecommand \href@noop [0]{\@secondoftwo}%
\providecommand \href [0]{\begingroup \@sanitize@url \@href}%
\providecommand \@href[1]{\@@startlink{#1}\@@href}%
\providecommand \@@href[1]{\endgroup#1\@@endlink}%
\providecommand \@sanitize@url [0]{\catcode `\\12\catcode `\$12\catcode
  `\&12\catcode `\#12\catcode `\^12\catcode `\_12\catcode `\%12\relax}%
\providecommand \@@startlink[1]{}%
\providecommand \@@endlink[0]{}%
\providecommand \url  [0]{\begingroup\@sanitize@url \@url }%
\providecommand \@url [1]{\endgroup\@href {#1}{\urlprefix }}%
\providecommand \urlprefix  [0]{URL }%
\providecommand \Eprint [0]{\href }%
\providecommand \doibase [0]{https://doi.org/}%
\providecommand \selectlanguage [0]{\@gobble}%
\providecommand \bibinfo  [0]{\@secondoftwo}%
\providecommand \bibfield  [0]{\@secondoftwo}%
\providecommand \translation [1]{[#1]}%
\providecommand \BibitemOpen [0]{}%
\providecommand \bibitemStop [0]{}%
\providecommand \bibitemNoStop [0]{.\EOS\space}%
\providecommand \EOS [0]{\spacefactor3000\relax}%
\providecommand \BibitemShut  [1]{\csname bibitem#1\endcsname}%
\let\auto@bib@innerbib\@empty
\bibitem [{\citenamefont {Aspelmeyer}\ \emph {et~al.}(2014)\citenamefont
  {Aspelmeyer}, \citenamefont {Kippenberg},\ and\ \citenamefont
  {Marquardt}}]{Aspelmeyer2014CavityOpt}%
  \BibitemOpen
  \bibfield  {author} {\bibinfo {author} {\bibfnamefont {M.}~\bibnamefont
  {Aspelmeyer}}, \bibinfo {author} {\bibfnamefont {T.~J.}\ \bibnamefont
  {Kippenberg}},\ and\ \bibinfo {author} {\bibfnamefont {F.}~\bibnamefont
  {Marquardt}},\ }\bibfield  {title} {\bibinfo {title} {Cavity
  {O}ptomechanics},\ }\href {https://doi.org/10.1103/RevModPhys.86.1391}
  {\bibfield  {journal} {\bibinfo  {journal} {Rev. Mod. Phys.}\ }\textbf
  {\bibinfo {volume} {86}},\ \bibinfo {pages} {1391} (\bibinfo {year}
  {2014})}\BibitemShut {NoStop}%
\bibitem [{\citenamefont {Davidovikj}\ \emph {et~al.}(2017)\citenamefont
  {Davidovikj}, \citenamefont {Alijani}, \citenamefont {Cartamil-Bueno},
  \citenamefont {van~der Zant}, \citenamefont {Amabili},\ and\ \citenamefont
  {Steeneken}}]{Davidovikj2017Nonlinear}%
  \BibitemOpen
  \bibfield  {author} {\bibinfo {author} {\bibfnamefont {D.}~\bibnamefont
  {Davidovikj}}, \bibinfo {author} {\bibfnamefont {F.}~\bibnamefont {Alijani}},
  \bibinfo {author} {\bibfnamefont {S.~J.}\ \bibnamefont {Cartamil-Bueno}},
  \bibinfo {author} {\bibfnamefont {H.~S.~J.}\ \bibnamefont {van~der Zant}},
  \bibinfo {author} {\bibfnamefont {M.}~\bibnamefont {Amabili}},\ and\ \bibinfo
  {author} {\bibfnamefont {P.~G.}\ \bibnamefont {Steeneken}},\ }\bibfield
  {title} {\bibinfo {title} {Nonlinear {D}ynamic {C}haracterization of
  {T}wo-{D}imensional {M}aterials},\ }\href
  {https://doi.org/10.1038/s41467-017-01351-4} {\bibfield  {journal} {\bibinfo
  {journal} {Nat. Commun.}\ }\textbf {\bibinfo {volume} {8}},\ \bibinfo {pages}
  {1253} (\bibinfo {year} {2017})}\BibitemShut {NoStop}%
\bibitem [{\citenamefont {Morell}\ \emph {et~al.}(2019)\citenamefont {Morell},
  \citenamefont {Tepsic}, \citenamefont {Reserbat-Plantey}, \citenamefont
  {Cepellotti}, \citenamefont {Manca}, \citenamefont {Epstein}, \citenamefont
  {Isacsson}, \citenamefont {Marie}, \citenamefont {Mauri},\ and\ \citenamefont
  {Bachtold}}]{Morell2019MoSe2}%
  \BibitemOpen
  \bibfield  {author} {\bibinfo {author} {\bibfnamefont {N.}~\bibnamefont
  {Morell}}, \bibinfo {author} {\bibfnamefont {S.}~\bibnamefont {Tepsic}},
  \bibinfo {author} {\bibfnamefont {A.}~\bibnamefont {Reserbat-Plantey}},
  \bibinfo {author} {\bibfnamefont {A.}~\bibnamefont {Cepellotti}}, \bibinfo
  {author} {\bibfnamefont {M.}~\bibnamefont {Manca}}, \bibinfo {author}
  {\bibfnamefont {I.}~\bibnamefont {Epstein}}, \bibinfo {author} {\bibfnamefont
  {A.}~\bibnamefont {Isacsson}}, \bibinfo {author} {\bibfnamefont
  {X.}~\bibnamefont {Marie}}, \bibinfo {author} {\bibfnamefont
  {F.}~\bibnamefont {Mauri}},\ and\ \bibinfo {author} {\bibfnamefont
  {A.}~\bibnamefont {Bachtold}},\ }\bibfield  {title} {\bibinfo {title}
  {Optomechanical {M}easurement of {T}hermal {T}ransport in {T}wo-{D}imensional
  {M}ose$_{2}$ {L}attices},\ }\bibfield  {journal} {\bibinfo  {journal} {Nano
  Lett.}\ }\href {https://doi.org/10.1021/acs.nanolett.9b00560}
  {10.1021/acs.nanolett.9b00560} (\bibinfo {year} {2019})\BibitemShut {NoStop}%
\bibitem [{\citenamefont {Waitz}\ \emph {et~al.}(2012)\citenamefont {Waitz},
  \citenamefont {Nößner}, \citenamefont {Hertkorn}, \citenamefont
  {Schecker},\ and\ \citenamefont {Scheer}}]{Waitz2012Spa}%
  \BibitemOpen
  \bibfield  {author} {\bibinfo {author} {\bibfnamefont {R.}~\bibnamefont
  {Waitz}}, \bibinfo {author} {\bibfnamefont {S.}~\bibnamefont {Nößner}},
  \bibinfo {author} {\bibfnamefont {M.}~\bibnamefont {Hertkorn}}, \bibinfo
  {author} {\bibfnamefont {O.}~\bibnamefont {Schecker}},\ and\ \bibinfo
  {author} {\bibfnamefont {E.}~\bibnamefont {Scheer}},\ }\bibfield  {title}
  {\bibinfo {title} {Mode {S}hape and {D}ispersion {R}elation of {B}ending
  {W}aves in {T}hin {S}ilicon {M}embranes},\ }\bibfield  {journal} {\bibinfo
  {journal} {Phys. Rev. B.}\ }\textbf {\bibinfo {volume} {85}},\ \href
  {https://doi.org/10.1103/PhysRevB.85.035324} {10.1103/PhysRevB.85.035324}
  (\bibinfo {year} {2012})\BibitemShut {NoStop}%
\bibitem [{\citenamefont {Wang}\ \emph {et~al.}(2014)\citenamefont {Wang},
  \citenamefont {Lee},\ and\ \citenamefont {Feng}}]{Wang2014Spectromicro}%
  \BibitemOpen
  \bibfield  {author} {\bibinfo {author} {\bibfnamefont {Z.}~\bibnamefont
  {Wang}}, \bibinfo {author} {\bibfnamefont {J.}~\bibnamefont {Lee}},\ and\
  \bibinfo {author} {\bibfnamefont {P.~X.}\ \bibnamefont {Feng}},\ }\bibfield
  {title} {\bibinfo {title} {Spatial {M}apping of {M}ultimode {B}rownian
  {M}otions in {H}igh-{F}requency {S}ilicon {C}arbide {M}icrodisk
  {R}esonators},\ }\href {https://doi.org/10.1038/ncomms6158} {\bibfield
  {journal} {\bibinfo  {journal} {Nat. Commun.}\ }\textbf {\bibinfo {volume}
  {5}},\ \bibinfo {pages} {5158} (\bibinfo {year} {2014})}\BibitemShut
  {NoStop}%
\bibitem [{\citenamefont {Davidovikj}\ \emph {et~al.}(2016)\citenamefont
  {Davidovikj}, \citenamefont {Slim}, \citenamefont {Cartamil-Bueno},
  \citenamefont {van~der Zant}, \citenamefont {Steeneken},\ and\ \citenamefont
  {Venstra}}]{Davidovikj2016Visual}%
  \BibitemOpen
  \bibfield  {author} {\bibinfo {author} {\bibfnamefont {D.}~\bibnamefont
  {Davidovikj}}, \bibinfo {author} {\bibfnamefont {J.~J.}\ \bibnamefont
  {Slim}}, \bibinfo {author} {\bibfnamefont {S.~J.}\ \bibnamefont
  {Cartamil-Bueno}}, \bibinfo {author} {\bibfnamefont {H.~S.}\ \bibnamefont
  {van~der Zant}}, \bibinfo {author} {\bibfnamefont {P.~G.}\ \bibnamefont
  {Steeneken}},\ and\ \bibinfo {author} {\bibfnamefont {W.~J.}\ \bibnamefont
  {Venstra}},\ }\bibfield  {title} {\bibinfo {title} {Visualizing the {M}otion
  of {G}raphene {N}anodrums},\ }\href
  {https://doi.org/10.1021/acs.nanolett.6b00477} {\bibfield  {journal}
  {\bibinfo  {journal} {Nano Lett.}\ }\textbf {\bibinfo {volume} {16}},\
  \bibinfo {pages} {2768} (\bibinfo {year} {2016})}\BibitemShut {NoStop}%
\bibitem [{\citenamefont {Kim}\ \emph {et~al.}(2018)\citenamefont {Kim},
  \citenamefont {Yu},\ and\ \citenamefont {van~der Zande}}]{Kim2018}%
  \BibitemOpen
  \bibfield  {author} {\bibinfo {author} {\bibfnamefont {S.}~\bibnamefont
  {Kim}}, \bibinfo {author} {\bibfnamefont {J.}~\bibnamefont {Yu}},\ and\
  \bibinfo {author} {\bibfnamefont {A.~M.}\ \bibnamefont {van~der Zande}},\
  }\bibfield  {title} {\bibinfo {title} {Nano-{E}lectromechanical {D}rumhead
  {R}esonators from {T}wo-{D}imensional {M}aterial {B}imorphs},\ }\href
  {https://doi.org/10.1021/acs.nanolett.8b01926} {\bibfield  {journal}
  {\bibinfo  {journal} {Nano Lett.}\ }\textbf {\bibinfo {volume} {18}},\
  \bibinfo {pages} {6686} (\bibinfo {year} {2018})}\BibitemShut {NoStop}%
\bibitem [{\citenamefont {Esmenda}\ \emph {et~al.}(2021)\citenamefont
  {Esmenda}, \citenamefont {Aguila}, \citenamefont {Wang}, \citenamefont {Lee},
  \citenamefont {Yang}, \citenamefont {Lin}, \citenamefont {Chang‐Liao},
  \citenamefont {Katz}, \citenamefont {Kafanov}, \citenamefont {Pashkin},\ and\
  \citenamefont {Chen}}]{esmenda:2020}%
  \BibitemOpen
  \bibfield  {author} {\bibinfo {author} {\bibfnamefont {J.~C.}\ \bibnamefont
  {Esmenda}}, \bibinfo {author} {\bibfnamefont {M.~A.~C.}\ \bibnamefont
  {Aguila}}, \bibinfo {author} {\bibfnamefont {J.}~\bibnamefont {Wang}},
  \bibinfo {author} {\bibfnamefont {T.}~\bibnamefont {Lee}}, \bibinfo {author}
  {\bibfnamefont {C.}~\bibnamefont {Yang}}, \bibinfo {author} {\bibfnamefont
  {K.}~\bibnamefont {Lin}}, \bibinfo {author} {\bibfnamefont {K.}~\bibnamefont
  {Chang‐Liao}}, \bibinfo {author} {\bibfnamefont {N.}~\bibnamefont {Katz}},
  \bibinfo {author} {\bibfnamefont {S.}~\bibnamefont {Kafanov}}, \bibinfo
  {author} {\bibfnamefont {Y.~A.}\ \bibnamefont {Pashkin}},\ and\ \bibinfo
  {author} {\bibfnamefont {C.}~\bibnamefont {Chen}},\ }\bibfield  {title}
  {\bibinfo {title} {Imaging {O}ff‐{R}esonance {N}anomechanical {M}otion as
  {M}odal {S}uperposition},\ }\bibfield  {journal} {\bibinfo  {journal} {Adv.
  Sci.}\ }\textbf {\bibinfo {volume} {8}},\ \href
  {https://doi.org/10.1002/advs.202005041} {10.1002/advs.202005041} (\bibinfo
  {year} {2021})\BibitemShut {NoStop}%
\bibitem [{\citenamefont {Purdy}\ \emph {et~al.}(2017)\citenamefont {Purdy},
  \citenamefont {Grutter}, \citenamefont {Srinivasan},\ and\ \citenamefont
  {Taylor}}]{Purdy2017QuantCorr}%
  \BibitemOpen
  \bibfield  {author} {\bibinfo {author} {\bibfnamefont {T.~P.}\ \bibnamefont
  {Purdy}}, \bibinfo {author} {\bibfnamefont {K.~E.}\ \bibnamefont {Grutter}},
  \bibinfo {author} {\bibfnamefont {K.}~\bibnamefont {Srinivasan}},\ and\
  \bibinfo {author} {\bibfnamefont {J.~M.}\ \bibnamefont {Taylor}},\ }\bibfield
   {title} {\bibinfo {title} {Quantum {C}orrelations from a
  {R}oom-{T}emperature {O}ptomechanical {C}avity},\ }\href
  {https://doi.org/10.1126/science.aag1407} {\bibfield  {journal} {\bibinfo
  {journal} {Science}\ }\textbf {\bibinfo {volume} {356}},\ \bibinfo {pages}
  {1265} (\bibinfo {year} {2017})}\BibitemShut {NoStop}%
\bibitem [{\citenamefont {Delic}\ \emph {et~al.}(2020)\citenamefont {Delic},
  \citenamefont {Reseinbauer}, \citenamefont {Dare}, \citenamefont {Grass},
  \citenamefont {Vuletic}, \citenamefont {Kiesel},\ and\ \citenamefont
  {Aspelmeyer}}]{Delic2020RTCooling}%
  \BibitemOpen
  \bibfield  {author} {\bibinfo {author} {\bibfnamefont {U.}~\bibnamefont
  {Delic}}, \bibinfo {author} {\bibfnamefont {M.}~\bibnamefont {Reseinbauer}},
  \bibinfo {author} {\bibfnamefont {K.}~\bibnamefont {Dare}}, \bibinfo {author}
  {\bibfnamefont {D.}~\bibnamefont {Grass}}, \bibinfo {author} {\bibfnamefont
  {V.}~\bibnamefont {Vuletic}}, \bibinfo {author} {\bibfnamefont
  {N.}~\bibnamefont {Kiesel}},\ and\ \bibinfo {author} {\bibfnamefont
  {M.}~\bibnamefont {Aspelmeyer}},\ }\bibfield  {title} {\bibinfo {title}
  {Cooling of a {L}evitated {N}anoparticle to the {M}otional {Q}uantum {G}round
  {S}tate},\ }\href {https://doi.org/10.1126/science.aba3993} {\bibfield
  {journal} {\bibinfo  {journal} {Science}\ }\textbf {\bibinfo {volume}
  {367}},\ \bibinfo {pages} {892} (\bibinfo {year} {2020})}\BibitemShut
  {NoStop}%
\bibitem [{\citenamefont {Bagci}\ \emph {et~al.}(2014)\citenamefont {Bagci},
  \citenamefont {Simonsen}, \citenamefont {Schmid}, \citenamefont {Villanueva},
  \citenamefont {Zeuthen}, \citenamefont {Appel}, \citenamefont {Taylor},
  \citenamefont {S{\o}rensen}, \citenamefont {Usami}, \citenamefont
  {Schliesser} \emph {et~al.}}]{bagci2014optical}%
  \BibitemOpen
  \bibfield  {author} {\bibinfo {author} {\bibfnamefont {T.}~\bibnamefont
  {Bagci}}, \bibinfo {author} {\bibfnamefont {A.}~\bibnamefont {Simonsen}},
  \bibinfo {author} {\bibfnamefont {S.}~\bibnamefont {Schmid}}, \bibinfo
  {author} {\bibfnamefont {L.~G.}\ \bibnamefont {Villanueva}}, \bibinfo
  {author} {\bibfnamefont {E.}~\bibnamefont {Zeuthen}}, \bibinfo {author}
  {\bibfnamefont {J.}~\bibnamefont {Appel}}, \bibinfo {author} {\bibfnamefont
  {J.~M.}\ \bibnamefont {Taylor}}, \bibinfo {author} {\bibfnamefont
  {A.}~\bibnamefont {S{\o}rensen}}, \bibinfo {author} {\bibfnamefont
  {K.}~\bibnamefont {Usami}}, \bibinfo {author} {\bibfnamefont
  {A.}~\bibnamefont {Schliesser}}, \emph {et~al.},\ }\bibfield  {title}
  {\bibinfo {title} {Optical {D}etection of {R}adio {W}aves {T}hrough a
  {N}anomechanical {T}ransducer},\ }\href {https://doi.org/10.1063/1.4862296}
  {\bibfield  {journal} {\bibinfo  {journal} {Nature}\ }\textbf {\bibinfo
  {volume} {507}},\ \bibinfo {pages} {81} (\bibinfo {year} {2014})}\BibitemShut
  {NoStop}%
\bibitem [{\citenamefont {Andrews}\ \emph {et~al.}(2014)\citenamefont
  {Andrews}, \citenamefont {Peterson}, \citenamefont {Purdy}, \citenamefont
  {Cicak}, \citenamefont {Simmonds}, \citenamefont {Regal},\ and\ \citenamefont
  {Lehnert}}]{andrews2014bidirectional}%
  \BibitemOpen
  \bibfield  {author} {\bibinfo {author} {\bibfnamefont {R.~W.}\ \bibnamefont
  {Andrews}}, \bibinfo {author} {\bibfnamefont {R.~W.}\ \bibnamefont
  {Peterson}}, \bibinfo {author} {\bibfnamefont {T.~P.}\ \bibnamefont {Purdy}},
  \bibinfo {author} {\bibfnamefont {K.}~\bibnamefont {Cicak}}, \bibinfo
  {author} {\bibfnamefont {R.~W.}\ \bibnamefont {Simmonds}}, \bibinfo {author}
  {\bibfnamefont {C.~A.}\ \bibnamefont {Regal}},\ and\ \bibinfo {author}
  {\bibfnamefont {K.~W.}\ \bibnamefont {Lehnert}},\ }\bibfield  {title}
  {\bibinfo {title} {Bidirectional and {E}fficient {C}onversion between
  {M}icrowave and {O}ptical {L}ight},\ }\href
  {https://doi.org/10.1038/nphys2911} {\bibfield  {journal} {\bibinfo
  {journal} {Nat. Phys.}\ }\textbf {\bibinfo {volume} {10}},\ \bibinfo {pages}
  {321} (\bibinfo {year} {2014})}\BibitemShut {NoStop}%
\bibitem [{\citenamefont {Higginbotham}\ \emph {et~al.}(2018)\citenamefont
  {Higginbotham}, \citenamefont {Burns}, \citenamefont {Urmey}, \citenamefont
  {Peterson}, \citenamefont {Kampel}, \citenamefont {Brubaker}, \citenamefont
  {Smith}, \citenamefont {Lehnert},\ and\ \citenamefont
  {Regal}}]{Higginbotham2018EO}%
  \BibitemOpen
  \bibfield  {author} {\bibinfo {author} {\bibfnamefont {A.~P.}\ \bibnamefont
  {Higginbotham}}, \bibinfo {author} {\bibfnamefont {P.~S.}\ \bibnamefont
  {Burns}}, \bibinfo {author} {\bibfnamefont {M.~D.}\ \bibnamefont {Urmey}},
  \bibinfo {author} {\bibfnamefont {R.~W.}\ \bibnamefont {Peterson}}, \bibinfo
  {author} {\bibfnamefont {N.~S.}\ \bibnamefont {Kampel}}, \bibinfo {author}
  {\bibfnamefont {B.~M.}\ \bibnamefont {Brubaker}}, \bibinfo {author}
  {\bibfnamefont {G.}~\bibnamefont {Smith}}, \bibinfo {author} {\bibfnamefont
  {K.~W.}\ \bibnamefont {Lehnert}},\ and\ \bibinfo {author} {\bibfnamefont
  {C.~A.}\ \bibnamefont {Regal}},\ }\bibfield  {title} {\bibinfo {title}
  {Harnessing {E}lectro-{O}ptic {C}orrelations in an {E}fficient {M}echanical
  {C}onverter},\ }\href {https://doi.org/10.1038/s41567-018-0210-0} {\bibfield
  {journal} {\bibinfo  {journal} {Nat. Phys.}\ }\textbf {\bibinfo {volume}
  {14}},\ \bibinfo {pages} {1038} (\bibinfo {year} {2018})}\BibitemShut
  {NoStop}%
\bibitem [{\citenamefont {Azak}\ \emph {et~al.}(2007)\citenamefont {Azak},
  \citenamefont {Shagam}, \citenamefont {Karabacak}, \citenamefont {Ekinci},
  \citenamefont {Kim},\ and\ \citenamefont {Jang}}]{Azak2007Fiber}%
  \BibitemOpen
  \bibfield  {author} {\bibinfo {author} {\bibfnamefont {N.~O.}\ \bibnamefont
  {Azak}}, \bibinfo {author} {\bibfnamefont {M.~Y.}\ \bibnamefont {Shagam}},
  \bibinfo {author} {\bibfnamefont {D.~M.}\ \bibnamefont {Karabacak}}, \bibinfo
  {author} {\bibfnamefont {K.~L.}\ \bibnamefont {Ekinci}}, \bibinfo {author}
  {\bibfnamefont {D.~H.}\ \bibnamefont {Kim}},\ and\ \bibinfo {author}
  {\bibfnamefont {D.~Y.}\ \bibnamefont {Jang}},\ }\bibfield  {title} {\bibinfo
  {title} {Nanomechanical {D}isplacement {D}etection {U}sing {F}iber-optic
  {I}nterferometry},\ }\href {https://doi.org/10.1063/1.2776981} {\bibfield
  {journal} {\bibinfo  {journal} {Appl. Phys. Lett.}\ }\textbf {\bibinfo
  {volume} {91}},\ \bibinfo {pages} {093112} (\bibinfo {year}
  {2007})}\BibitemShut {NoStop}%
\bibitem [{\citenamefont {Metzger}\ \emph {et~al.}(2008)\citenamefont
  {Metzger}, \citenamefont {Favero}, \citenamefont {Ortlieb},\ and\
  \citenamefont {Karrai}}]{Metzger2008pth}%
  \BibitemOpen
  \bibfield  {author} {\bibinfo {author} {\bibfnamefont {C.}~\bibnamefont
  {Metzger}}, \bibinfo {author} {\bibfnamefont {I.}~\bibnamefont {Favero}},
  \bibinfo {author} {\bibfnamefont {A.}~\bibnamefont {Ortlieb}},\ and\ \bibinfo
  {author} {\bibfnamefont {K.}~\bibnamefont {Karrai}},\ }\bibfield  {title}
  {\bibinfo {title} {Optical {S}elf {C}ooling of a {D}eformable {F}abry-{P}erot
  {C}avity in the {C}lassical {L}imit},\ }\bibfield  {journal} {\bibinfo
  {journal} {Phys. Rev. B.}\ }\textbf {\bibinfo {volume} {78}},\ \href
  {https://doi.org/10.1103/PhysRevB.78.035309} {10.1103/PhysRevB.78.035309}
  (\bibinfo {year} {2008})\BibitemShut {NoStop}%
\bibitem [{\citenamefont {Flowers-Jacobs}\ \emph {et~al.}(2012)\citenamefont
  {Flowers-Jacobs}, \citenamefont {Hoch}, \citenamefont {Sankey}, \citenamefont
  {Kashkanova}, \citenamefont {Jayich}, \citenamefont {Deutsch}, \citenamefont
  {Reichel},\ and\ \citenamefont {Harris}}]{FJ2012fpopt}%
  \BibitemOpen
  \bibfield  {author} {\bibinfo {author} {\bibfnamefont {N.~E.}\ \bibnamefont
  {Flowers-Jacobs}}, \bibinfo {author} {\bibfnamefont {S.~W.}\ \bibnamefont
  {Hoch}}, \bibinfo {author} {\bibfnamefont {J.~C.}\ \bibnamefont {Sankey}},
  \bibinfo {author} {\bibfnamefont {A.}~\bibnamefont {Kashkanova}}, \bibinfo
  {author} {\bibfnamefont {A.~M.}\ \bibnamefont {Jayich}}, \bibinfo {author}
  {\bibfnamefont {C.}~\bibnamefont {Deutsch}}, \bibinfo {author} {\bibfnamefont
  {J.}~\bibnamefont {Reichel}},\ and\ \bibinfo {author} {\bibfnamefont
  {J.~G.~E.}\ \bibnamefont {Harris}},\ }\bibfield  {title} {\bibinfo {title}
  {Fiber-{C}avity-{B}ased {O}ptomechanical {D}evice},\ }\bibfield  {journal}
  {\bibinfo  {journal} {Appl. Phys. Lett.}\ }\textbf {\bibinfo {volume}
  {101}},\ \href {https://doi.org/10.1063/1.4768779} {10.1063/1.4768779}
  (\bibinfo {year} {2012})\BibitemShut {NoStop}%
\bibitem [{\citenamefont {Basarir}\ \emph {et~al.}(2012)\citenamefont
  {Basarir}, \citenamefont {Bramhavar},\ and\ \citenamefont
  {Ekinci}}]{Basarir2012MotionTrans}%
  \BibitemOpen
  \bibfield  {author} {\bibinfo {author} {\bibfnamefont {O.}~\bibnamefont
  {Basarir}}, \bibinfo {author} {\bibfnamefont {S.}~\bibnamefont {Bramhavar}},\
  and\ \bibinfo {author} {\bibfnamefont {K.~L.}\ \bibnamefont {Ekinci}},\
  }\bibfield  {title} {\bibinfo {title} {Motion {T}ransduction in
  {N}anoelectromechanical {S}ystems ({NEMS}) {A}rrays {U}sing {N}ear-{F}ield
  {O}ptomechanical {C}oupling},\ }\href {https://doi.org/10.1021/nl2031585}
  {\bibfield  {journal} {\bibinfo  {journal} {Nano Lett.}\ }\textbf {\bibinfo
  {volume} {12}},\ \bibinfo {pages} {534} (\bibinfo {year} {2012})}\BibitemShut
  {NoStop}%
\bibitem [{\citenamefont {Vainsencher}\ \emph {et~al.}(2016)\citenamefont
  {Vainsencher}, \citenamefont {Satzinger}, \citenamefont {Peairs},\ and\
  \citenamefont {Cleland}}]{Vainsencher2016Bidirec}%
  \BibitemOpen
  \bibfield  {author} {\bibinfo {author} {\bibfnamefont {A.}~\bibnamefont
  {Vainsencher}}, \bibinfo {author} {\bibfnamefont {K.~J.}\ \bibnamefont
  {Satzinger}}, \bibinfo {author} {\bibfnamefont {G.~A.}\ \bibnamefont
  {Peairs}},\ and\ \bibinfo {author} {\bibfnamefont {A.~N.}\ \bibnamefont
  {Cleland}},\ }\bibfield  {title} {\bibinfo {title} {Bi-{D}irectional
  {C}onversion {B}etween {M}icrowave and {O}ptical {F}requencies in a
  {P}iezoelectric {O}ptomechanical {D}evice},\ }\bibfield  {journal} {\bibinfo
  {journal} {Appl. Phys. Lett.}\ }\textbf {\bibinfo {volume} {109}},\ \href
  {https://doi.org/10.1063/1.4955408} {10.1063/1.4955408} (\bibinfo {year}
  {2016})\BibitemShut {NoStop}%
\bibitem [{\citenamefont {Midolo}\ \emph {et~al.}(2018)\citenamefont {Midolo},
  \citenamefont {Schliesser},\ and\ \citenamefont {Fiore}}]{Midolo2018NOEMS}%
  \BibitemOpen
  \bibfield  {author} {\bibinfo {author} {\bibfnamefont {L.}~\bibnamefont
  {Midolo}}, \bibinfo {author} {\bibfnamefont {A.}~\bibnamefont {Schliesser}},\
  and\ \bibinfo {author} {\bibfnamefont {A.}~\bibnamefont {Fiore}},\ }\bibfield
   {title} {\bibinfo {title} {Nano-{O}pto-{E}lectro-{M}echanical {S}ystems},\
  }\href {https://doi.org/10.1038/s41565-017-0039-1} {\bibfield  {journal}
  {\bibinfo  {journal} {Nat. Nanotechnol.}\ }\textbf {\bibinfo {volume} {13}},\
  \bibinfo {pages} {11} (\bibinfo {year} {2018})}\BibitemShut {NoStop}%
\bibitem [{\citenamefont {Chen}\ \emph {et~al.}(2018)\citenamefont {Chen},
  \citenamefont {Yang}, \citenamefont {Mao}, \citenamefont {Lu}, \citenamefont
  {Sch{\"a}dler}, \citenamefont {Reserbat-Plantey}, \citenamefont {Osmond},
  \citenamefont {Cao}, \citenamefont {Li}, \citenamefont {Wang} \emph
  {et~al.}}]{chen2018vibration}%
  \BibitemOpen
  \bibfield  {author} {\bibinfo {author} {\bibfnamefont {F.}~\bibnamefont
  {Chen}}, \bibinfo {author} {\bibfnamefont {C.}~\bibnamefont {Yang}}, \bibinfo
  {author} {\bibfnamefont {W.}~\bibnamefont {Mao}}, \bibinfo {author}
  {\bibfnamefont {H.}~\bibnamefont {Lu}}, \bibinfo {author} {\bibfnamefont
  {K.~G.}\ \bibnamefont {Sch{\"a}dler}}, \bibinfo {author} {\bibfnamefont
  {A.}~\bibnamefont {Reserbat-Plantey}}, \bibinfo {author} {\bibfnamefont
  {J.}~\bibnamefont {Osmond}}, \bibinfo {author} {\bibfnamefont
  {G.}~\bibnamefont {Cao}}, \bibinfo {author} {\bibfnamefont {X.}~\bibnamefont
  {Li}}, \bibinfo {author} {\bibfnamefont {C.}~\bibnamefont {Wang}}, \emph
  {et~al.},\ }\bibfield  {title} {\bibinfo {title} {Vibration {D}etection
  {S}chemes {B}ased on {A}bsorbance {T}uning in {M}onolayer {M}olybdenum
  {D}isulfide {M}echanical {R}esonators},\ }\href
  {https://doi.org/10.1088/2053-1583/aae5b7} {\bibfield  {journal} {\bibinfo
  {journal} {2D Mater.}\ }\textbf {\bibinfo {volume} {6}},\ \bibinfo {pages}
  {011003} (\bibinfo {year} {2018})}\BibitemShut {NoStop}%
\bibitem [{\citenamefont {Barton}\ \emph {et~al.}(2012)\citenamefont {Barton},
  \citenamefont {Storch}, \citenamefont {Adiga}, \citenamefont {Sakakibara},
  \citenamefont {Cipriany}, \citenamefont {Ilic}, \citenamefont {Wang},
  \citenamefont {Ong}, \citenamefont {McEuen}, \citenamefont {Parpia},\ and\
  \citenamefont {Craighead}}]{Barton2012pth}%
  \BibitemOpen
  \bibfield  {author} {\bibinfo {author} {\bibfnamefont {R.~A.}\ \bibnamefont
  {Barton}}, \bibinfo {author} {\bibfnamefont {I.~R.}\ \bibnamefont {Storch}},
  \bibinfo {author} {\bibfnamefont {V.~P.}\ \bibnamefont {Adiga}}, \bibinfo
  {author} {\bibfnamefont {R.}~\bibnamefont {Sakakibara}}, \bibinfo {author}
  {\bibfnamefont {B.~R.}\ \bibnamefont {Cipriany}}, \bibinfo {author}
  {\bibfnamefont {B.}~\bibnamefont {Ilic}}, \bibinfo {author} {\bibfnamefont
  {S.~P.}\ \bibnamefont {Wang}}, \bibinfo {author} {\bibfnamefont
  {P.}~\bibnamefont {Ong}}, \bibinfo {author} {\bibfnamefont {P.~L.}\
  \bibnamefont {McEuen}}, \bibinfo {author} {\bibfnamefont {J.~M.}\
  \bibnamefont {Parpia}},\ and\ \bibinfo {author} {\bibfnamefont {H.~G.}\
  \bibnamefont {Craighead}},\ }\bibfield  {title} {\bibinfo {title}
  {Photothermal {S}elf-{O}scillation and {L}aser {C}ooling of {G}raphene
  {O}ptomechanical {S}ystems},\ }\href {https://doi.org/10.1021/nl302036x}
  {\bibfield  {journal} {\bibinfo  {journal} {Nano Lett.}\ }\textbf {\bibinfo
  {volume} {12}},\ \bibinfo {pages} {4681} (\bibinfo {year}
  {2012})}\BibitemShut {NoStop}%
\bibitem [{\citenamefont {Primo}\ \emph {et~al.}(2021)\citenamefont {Primo},
  \citenamefont {Kersul}, \citenamefont {Benevides}, \citenamefont {Carvalho},
  \citenamefont {Ménard}, \citenamefont {Frateschi}, \citenamefont {de~Assis},
  \citenamefont {Wiederhecker},\ and\ \citenamefont
  {Mayer~Alegre}}]{Primo2021pthmodelling}%
  \BibitemOpen
  \bibfield  {author} {\bibinfo {author} {\bibfnamefont {A.~G.}\ \bibnamefont
  {Primo}}, \bibinfo {author} {\bibfnamefont {C.~M.}\ \bibnamefont {Kersul}},
  \bibinfo {author} {\bibfnamefont {R.}~\bibnamefont {Benevides}}, \bibinfo
  {author} {\bibfnamefont {N.~C.}\ \bibnamefont {Carvalho}}, \bibinfo {author}
  {\bibfnamefont {M.}~\bibnamefont {Ménard}}, \bibinfo {author} {\bibfnamefont
  {N.~C.}\ \bibnamefont {Frateschi}}, \bibinfo {author} {\bibfnamefont {P.-L.}\
  \bibnamefont {de~Assis}}, \bibinfo {author} {\bibfnamefont {G.~S.}\
  \bibnamefont {Wiederhecker}},\ and\ \bibinfo {author} {\bibfnamefont {T.~P.}\
  \bibnamefont {Mayer~Alegre}},\ }\bibfield  {title} {\bibinfo {title}
  {Accurate {M}odeling and {C}haracterization of {P}hotothermal {F}orces in
  {O}ptomechanics},\ }\bibfield  {journal} {\bibinfo  {journal} {APL
  Photonics}\ }\textbf {\bibinfo {volume} {6}},\ \href
  {https://doi.org/10.1063/5.0055201} {10.1063/5.0055201} (\bibinfo {year}
  {2021})\BibitemShut {NoStop}%
\bibitem [{\citenamefont {Kippenberg}\ and\ \citenamefont
  {Vahala}(2008)}]{Kippenberg2008CavFeed}%
  \BibitemOpen
  \bibfield  {author} {\bibinfo {author} {\bibfnamefont {T.~J.}\ \bibnamefont
  {Kippenberg}}\ and\ \bibinfo {author} {\bibfnamefont {K.~J.}\ \bibnamefont
  {Vahala}},\ }\bibfield  {title} {\bibinfo {title} {Cavity {O}ptomechanics:
  {B}ack at the {M}esoscale},\ }\href {https://doi.org/10.1126/science.1156032}
  {\bibfield  {journal} {\bibinfo  {journal} {Science}\ }\textbf {\bibinfo
  {volume} {321}},\ \bibinfo {pages} {1172} (\bibinfo {year}
  {2008})}\BibitemShut {NoStop}%
\bibitem [{\citenamefont {Barton}\ \emph {et~al.}(2011)\citenamefont {Barton},
  \citenamefont {Ilic}, \citenamefont {van~der Zande}, \citenamefont {Whitney},
  \citenamefont {McEuen}, \citenamefont {Parpia},\ and\ \citenamefont
  {Craighead}}]{Barton2011Q}%
  \BibitemOpen
  \bibfield  {author} {\bibinfo {author} {\bibfnamefont {R.~A.}\ \bibnamefont
  {Barton}}, \bibinfo {author} {\bibfnamefont {B.}~\bibnamefont {Ilic}},
  \bibinfo {author} {\bibfnamefont {A.~M.}\ \bibnamefont {van~der Zande}},
  \bibinfo {author} {\bibfnamefont {W.~S.}\ \bibnamefont {Whitney}}, \bibinfo
  {author} {\bibfnamefont {P.~L.}\ \bibnamefont {McEuen}}, \bibinfo {author}
  {\bibfnamefont {J.~M.}\ \bibnamefont {Parpia}},\ and\ \bibinfo {author}
  {\bibfnamefont {H.~G.}\ \bibnamefont {Craighead}},\ }\bibfield  {title}
  {\bibinfo {title} {High, {S}ize-{D}ependent {Q}uality {F}actor in an {A}rray
  of {G}raphene {M}echanical {R}esonators},\ }\href
  {https://doi.org/10.1021/nl1042227} {\bibfield  {journal} {\bibinfo
  {journal} {Nano Lett.}\ }\textbf {\bibinfo {volume} {11}},\ \bibinfo {pages}
  {1232} (\bibinfo {year} {2011})}\BibitemShut {NoStop}%
\bibitem [{\citenamefont {Lemme}\ \emph {et~al.}(2020)\citenamefont {Lemme},
  \citenamefont {Wagner}, \citenamefont {Lee}, \citenamefont {Fan},
  \citenamefont {Verbiest}, \citenamefont {Wittmann}, \citenamefont {Lukas},
  \citenamefont {Dolleman}, \citenamefont {Niklaus}, \citenamefont {van~der
  Zant}, \citenamefont {Duesberg},\ and\ \citenamefont
  {Steeneken}}]{Lemme2020NEMS}%
  \BibitemOpen
  \bibfield  {author} {\bibinfo {author} {\bibfnamefont {M.~C.}\ \bibnamefont
  {Lemme}}, \bibinfo {author} {\bibfnamefont {S.}~\bibnamefont {Wagner}},
  \bibinfo {author} {\bibfnamefont {K.}~\bibnamefont {Lee}}, \bibinfo {author}
  {\bibfnamefont {X.}~\bibnamefont {Fan}}, \bibinfo {author} {\bibfnamefont
  {G.~J.}\ \bibnamefont {Verbiest}}, \bibinfo {author} {\bibfnamefont
  {S.}~\bibnamefont {Wittmann}}, \bibinfo {author} {\bibfnamefont
  {S.}~\bibnamefont {Lukas}}, \bibinfo {author} {\bibfnamefont {R.~J.}\
  \bibnamefont {Dolleman}}, \bibinfo {author} {\bibfnamefont {F.}~\bibnamefont
  {Niklaus}}, \bibinfo {author} {\bibfnamefont {H.~S.~J.}\ \bibnamefont
  {van~der Zant}}, \bibinfo {author} {\bibfnamefont {G.~S.}\ \bibnamefont
  {Duesberg}},\ and\ \bibinfo {author} {\bibfnamefont {P.~G.}\ \bibnamefont
  {Steeneken}},\ }\bibfield  {title} {\bibinfo {title} {Nanoelectromechanical
  {S}ensors {B}ased on {S}uspended 2{D} {M}aterials},\ }\href
  {https://doi.org/10.34133/2020/8748602} {\bibfield  {journal} {\bibinfo
  {journal} {Research}\ }\textbf {\bibinfo {volume} {2020}},\ \bibinfo {pages}
  {8748602} (\bibinfo {year} {2020})}\BibitemShut {NoStop}%
\bibitem [{\citenamefont {Steeneken}\ \emph {et~al.}(2021)\citenamefont
  {Steeneken}, \citenamefont {Dolleman}, \citenamefont {Davidovikj},
  \citenamefont {Alijani},\ and\ \citenamefont {van~der
  Zant}}]{Steeneken2021Dynamics}%
  \BibitemOpen
  \bibfield  {author} {\bibinfo {author} {\bibfnamefont {P.~G.}\ \bibnamefont
  {Steeneken}}, \bibinfo {author} {\bibfnamefont {R.~J.}\ \bibnamefont
  {Dolleman}}, \bibinfo {author} {\bibfnamefont {D.}~\bibnamefont
  {Davidovikj}}, \bibinfo {author} {\bibfnamefont {F.}~\bibnamefont
  {Alijani}},\ and\ \bibinfo {author} {\bibfnamefont {H.~S.~J.}\ \bibnamefont
  {van~der Zant}},\ }\bibfield  {title} {\bibinfo {title} {Dynamics of 2{D}
  {M}aterial {M}embranes},\ }\bibfield  {journal} {\bibinfo  {journal} {2D
  Mater.}\ }\textbf {\bibinfo {volume} {8}},\ \href
  {https://doi.org/10.1088/2053-1583/ac152c} {10.1088/2053-1583/ac152c}
  (\bibinfo {year} {2021})\BibitemShut {NoStop}%
\bibitem [{\citenamefont {Blaikie}\ \emph {et~al.}(2019)\citenamefont
  {Blaikie}, \citenamefont {Miller},\ and\ \citenamefont
  {Aleman}}]{Blaikie2019GrBol}%
  \BibitemOpen
  \bibfield  {author} {\bibinfo {author} {\bibfnamefont {A.}~\bibnamefont
  {Blaikie}}, \bibinfo {author} {\bibfnamefont {D.}~\bibnamefont {Miller}},\
  and\ \bibinfo {author} {\bibfnamefont {B.~J.}\ \bibnamefont {Aleman}},\
  }\bibfield  {title} {\bibinfo {title} {A {F}ast and {S}ensitive
  {R}oom-{T}emperature {G}raphene {N}anomechanical {B}olometer},\ }\href
  {https://doi.org/10.1038/s41467-019-12562-2} {\bibfield  {journal} {\bibinfo
  {journal} {Nat. Commun.}\ }\textbf {\bibinfo {volume} {10}},\ \bibinfo
  {pages} {4726} (\bibinfo {year} {2019})}\BibitemShut {NoStop}%
\bibitem [{\citenamefont {Hu}\ \emph {et~al.}(2020)\citenamefont {Hu},
  \citenamefont {Xu}, \citenamefont {Xiang}, \citenamefont {Chen},
  \citenamefont {Zhou}, \citenamefont {Wang}, \citenamefont {Guo},
  \citenamefont {Ruan}, \citenamefont {Hu}, \citenamefont {Li}, \citenamefont
  {Liang}, \citenamefont {Jiang},\ and\ \citenamefont {Li}}]{Hu2020UVphoto}%
  \BibitemOpen
  \bibfield  {author} {\bibinfo {author} {\bibfnamefont {X.}~\bibnamefont
  {Hu}}, \bibinfo {author} {\bibfnamefont {E.}~\bibnamefont {Xu}}, \bibinfo
  {author} {\bibfnamefont {S.}~\bibnamefont {Xiang}}, \bibinfo {author}
  {\bibfnamefont {Z.}~\bibnamefont {Chen}}, \bibinfo {author} {\bibfnamefont
  {X.}~\bibnamefont {Zhou}}, \bibinfo {author} {\bibfnamefont {N.}~\bibnamefont
  {Wang}}, \bibinfo {author} {\bibfnamefont {H.}~\bibnamefont {Guo}}, \bibinfo
  {author} {\bibfnamefont {L.}~\bibnamefont {Ruan}}, \bibinfo {author}
  {\bibfnamefont {Y.}~\bibnamefont {Hu}}, \bibinfo {author} {\bibfnamefont
  {C.}~\bibnamefont {Li}}, \bibinfo {author} {\bibfnamefont {D.}~\bibnamefont
  {Liang}}, \bibinfo {author} {\bibfnamefont {Y.}~\bibnamefont {Jiang}},\ and\
  \bibinfo {author} {\bibfnamefont {G.}~\bibnamefont {Li}},\ }\bibfield
  {title} {\bibinfo {title} {Synthesis of {N}b{S}e$_{2}$ {S}ingle-{C}rystalline
  {N}anosheet {A}rrays for {UV} {P}hotodetectors},\ }\href
  {https://doi.org/10.1039/d0ce01140a} {\bibfield  {journal} {\bibinfo
  {journal} {CrystEngComm}\ }\textbf {\bibinfo {volume} {22}},\ \bibinfo
  {pages} {5710} (\bibinfo {year} {2020})}\BibitemShut {NoStop}%
\bibitem [{\citenamefont {Orchin}\ \emph {et~al.}(2019)\citenamefont {Orchin},
  \citenamefont {De~Fazio}, \citenamefont {Di~Bernardo}, \citenamefont {Hamer},
  \citenamefont {Yoon}, \citenamefont {Cadore}, \citenamefont {Goykhman},
  \citenamefont {Watanabe}, \citenamefont {Taniguchi}, \citenamefont
  {Robinson}, \citenamefont {Gorbachev}, \citenamefont {Ferrari},\ and\
  \citenamefont {Hadfield}}]{Orchin2019}%
  \BibitemOpen
  \bibfield  {author} {\bibinfo {author} {\bibfnamefont {G.~J.}\ \bibnamefont
  {Orchin}}, \bibinfo {author} {\bibfnamefont {D.}~\bibnamefont {De~Fazio}},
  \bibinfo {author} {\bibfnamefont {A.}~\bibnamefont {Di~Bernardo}}, \bibinfo
  {author} {\bibfnamefont {M.}~\bibnamefont {Hamer}}, \bibinfo {author}
  {\bibfnamefont {D.}~\bibnamefont {Yoon}}, \bibinfo {author} {\bibfnamefont
  {A.~R.}\ \bibnamefont {Cadore}}, \bibinfo {author} {\bibfnamefont
  {I.}~\bibnamefont {Goykhman}}, \bibinfo {author} {\bibfnamefont
  {K.}~\bibnamefont {Watanabe}}, \bibinfo {author} {\bibfnamefont
  {T.}~\bibnamefont {Taniguchi}}, \bibinfo {author} {\bibfnamefont {J.~W.~A.}\
  \bibnamefont {Robinson}}, \bibinfo {author} {\bibfnamefont {R.~V.}\
  \bibnamefont {Gorbachev}}, \bibinfo {author} {\bibfnamefont {A.~C.}\
  \bibnamefont {Ferrari}},\ and\ \bibinfo {author} {\bibfnamefont {R.~H.}\
  \bibnamefont {Hadfield}},\ }\bibfield  {title} {\bibinfo {title} {Niobium
  {D}iselenide {S}uperconducting {P}hotodetectors},\ }\href
  {https://doi.org/10.1063/1.5097389} {\bibfield  {journal} {\bibinfo
  {journal} {Appl. Phys. Lett.}\ }\textbf {\bibinfo {volume} {114}},\ \bibinfo
  {pages} {1} (\bibinfo {year} {2019})}\BibitemShut {NoStop}%
\bibitem [{\citenamefont {Jin}\ \emph {et~al.}(2021)\citenamefont {Jin},
  \citenamefont {Ji}, \citenamefont {Gu}, \citenamefont {Xie}, \citenamefont
  {Zhang}, \citenamefont {Wu},\ and\ \citenamefont {Cai}}]{Jin2021CryoNbSe2}%
  \BibitemOpen
  \bibfield  {author} {\bibinfo {author} {\bibfnamefont {Y.}~\bibnamefont
  {Jin}}, \bibinfo {author} {\bibfnamefont {Z.}~\bibnamefont {Ji}}, \bibinfo
  {author} {\bibfnamefont {F.}~\bibnamefont {Gu}}, \bibinfo {author}
  {\bibfnamefont {B.}~\bibnamefont {Xie}}, \bibinfo {author} {\bibfnamefont
  {R.}~\bibnamefont {Zhang}}, \bibinfo {author} {\bibfnamefont
  {J.}~\bibnamefont {Wu}},\ and\ \bibinfo {author} {\bibfnamefont
  {X.}~\bibnamefont {Cai}},\ }\bibfield  {title} {\bibinfo {title} {Multiple
  {M}echanisms of the {L}ow {T}emperature {P}hotoresponse in {N}iobium
  {D}iselenide},\ }\bibfield  {journal} {\bibinfo  {journal} {Appl. Phys.
  Lett.}\ }\textbf {\bibinfo {volume} {119}},\ \href
  {https://doi.org/10.1063/5.0073605} {10.1063/5.0073605} (\bibinfo {year}
  {2021})\BibitemShut {NoStop}%
\bibitem [{\citenamefont {Papas}\ \emph {et~al.}(2021)\citenamefont {Papas},
  \citenamefont {Ou}, \citenamefont {Plum},\ and\ \citenamefont
  {Zheludev}}]{Papas2021omn}%
  \BibitemOpen
  \bibfield  {author} {\bibinfo {author} {\bibfnamefont {D.}~\bibnamefont
  {Papas}}, \bibinfo {author} {\bibfnamefont {J.-Y.}\ \bibnamefont {Ou}},
  \bibinfo {author} {\bibfnamefont {E.}~\bibnamefont {Plum}},\ and\ \bibinfo
  {author} {\bibfnamefont {N.~I.}\ \bibnamefont {Zheludev}},\ }\bibfield
  {title} {\bibinfo {title} {Optomechanical {M}etamaterial {N}anobolometer},\
  }\bibfield  {journal} {\bibinfo  {journal} {APL Photonics}\ }\textbf
  {\bibinfo {volume} {6}},\ \href {https://doi.org/10.1063/5.0073583}
  {10.1063/5.0073583} (\bibinfo {year} {2021})\BibitemShut {NoStop}%
\bibitem [{\citenamefont {Ferreiro}(2019)}]{Ferreiro2019}%
  \BibitemOpen
  \bibfield  {author} {\bibinfo {author} {\bibfnamefont {J.~M.}\ \bibnamefont
  {Ferreiro}},\ }\emph {\bibinfo {title} {Experimental and {S}imulation {S}tudy
  of {E}lectron and {P}honon {P}roperties in {C}rystalline {M}aterials}},\
  \href {https://curate.nd.edu/show/vq27zk55530} {\bibinfo {type}
  {Dissertation}},\ \bibinfo  {school} {University of Notre Dame} (\bibinfo
  {year} {2019})\BibitemShut {NoStop}%
\bibitem [{\citenamefont {Sengupta}\ \emph {et~al.}(2010)\citenamefont
  {Sengupta}, \citenamefont {Solanki}, \citenamefont {Singh}, \citenamefont
  {Dhara},\ and\ \citenamefont {Deshmukh}}]{sengupta2010electromechanical}%
  \BibitemOpen
  \bibfield  {author} {\bibinfo {author} {\bibfnamefont {S.}~\bibnamefont
  {Sengupta}}, \bibinfo {author} {\bibfnamefont {H.~S.}\ \bibnamefont
  {Solanki}}, \bibinfo {author} {\bibfnamefont {V.}~\bibnamefont {Singh}},
  \bibinfo {author} {\bibfnamefont {S.}~\bibnamefont {Dhara}},\ and\ \bibinfo
  {author} {\bibfnamefont {M.~M.}\ \bibnamefont {Deshmukh}},\ }\bibfield
  {title} {\bibinfo {title} {Electromechanical {R}esonators as {P}robes of the
  {C}harge {D}ensity {W}ave {T}ransition at the {N}anoscale in {NbSe}$_{2}$},\
  }\href {https://doi.org/10.1103/PhysRevB.82.155432} {\bibfield  {journal}
  {\bibinfo  {journal} {Phys. Rev. B.}\ }\textbf {\bibinfo {volume} {82}},\
  \bibinfo {pages} {155432} (\bibinfo {year} {2010})}\BibitemShut {NoStop}%
\bibitem [{\citenamefont {Sun}\ \emph {et~al.}(2021)\citenamefont {Sun},
  \citenamefont {Agrawal},\ and\ \citenamefont {Singh}}]{Sun2021NbSe2}%
  \BibitemOpen
  \bibfield  {author} {\bibinfo {author} {\bibfnamefont {H.}~\bibnamefont
  {Sun}}, \bibinfo {author} {\bibfnamefont {P.}~\bibnamefont {Agrawal}},\ and\
  \bibinfo {author} {\bibfnamefont {C.~V.}\ \bibnamefont {Singh}},\ }\bibfield
  {title} {\bibinfo {title} {A {F}irst-{P}rinciples {S}tudy of the
  {R}elationship {B}etween {M}odulus and {I}deal {S}trength of
  {S}ingle-{L}ayer, {T}ransition {M}etal {D}ichalcogenides},\ }\href
  {https://doi.org/10.1039/d1ma00239b} {\bibfield  {journal} {\bibinfo
  {journal} {Materials Advances}\ }\textbf {\bibinfo {volume} {2}},\ \bibinfo
  {pages} {6631} (\bibinfo {year} {2021})}\BibitemShut {NoStop}%
\bibitem [{\citenamefont {Aguila}\ \emph {et~al.}(2022)\citenamefont {Aguila},
  \citenamefont {Esmenda}, \citenamefont {Wang}, \citenamefont {Lee},
  \citenamefont {Yang}, \citenamefont {Lin}, \citenamefont {Chang-Liao},
  \citenamefont {Kafanov}, \citenamefont {Pashkin},\ and\ \citenamefont
  {Chen}}]{Aguila2022FPCal}%
  \BibitemOpen
  \bibfield  {author} {\bibinfo {author} {\bibfnamefont {M.~A.~C.}\
  \bibnamefont {Aguila}}, \bibinfo {author} {\bibfnamefont {J.~C.}\
  \bibnamefont {Esmenda}}, \bibinfo {author} {\bibfnamefont {J.-Y.}\
  \bibnamefont {Wang}}, \bibinfo {author} {\bibfnamefont {T.-H.}\ \bibnamefont
  {Lee}}, \bibinfo {author} {\bibfnamefont {C.-Y.}\ \bibnamefont {Yang}},
  \bibinfo {author} {\bibfnamefont {K.-H.}\ \bibnamefont {Lin}}, \bibinfo
  {author} {\bibfnamefont {K.-S.}\ \bibnamefont {Chang-Liao}}, \bibinfo
  {author} {\bibfnamefont {S.}~\bibnamefont {Kafanov}}, \bibinfo {author}
  {\bibfnamefont {Y.~A.}\ \bibnamefont {Pashkin}},\ and\ \bibinfo {author}
  {\bibfnamefont {C.-D.}\ \bibnamefont {Chen}},\ }\bibfield  {title} {\bibinfo
  {title} {Fabry–{P}erot {I}nterferometric {C}alibration of {V}an der {W}aals
  {M}aterial-{B}ased {N}anomechanical {R}esonators},\ }\bibfield  {journal}
  {\bibinfo  {journal} {Nanoscale Adv.}\ }\href
  {https://doi.org/10.1039/d1na00794g} {10.1039/d1na00794g} (\bibinfo {year}
  {2022})\BibitemShut {NoStop}%
\bibitem [{\citenamefont {Lee}\ \emph {et~al.}(2013)\citenamefont {Lee},
  \citenamefont {Wang}, \citenamefont {He}, \citenamefont {Shan},\ and\
  \citenamefont {Feng}}]{Lee2013}%
  \BibitemOpen
  \bibfield  {author} {\bibinfo {author} {\bibfnamefont {J.}~\bibnamefont
  {Lee}}, \bibinfo {author} {\bibfnamefont {Z.}~\bibnamefont {Wang}}, \bibinfo
  {author} {\bibfnamefont {K.}~\bibnamefont {He}}, \bibinfo {author}
  {\bibfnamefont {J.}~\bibnamefont {Shan}},\ and\ \bibinfo {author}
  {\bibfnamefont {P.~X.-L.}\ \bibnamefont {Feng}},\ }\bibfield  {title}
  {\bibinfo {title} {High {F}requency {M}o{S}$_{2}$ {N}anomechanical
  {R}esonators},\ }\href {https://doi.org/10.1021/nn4018872} {\bibfield
  {journal} {\bibinfo  {journal} {ACS Nano}\ }\textbf {\bibinfo {volume} {7}},\
  \bibinfo {pages} {6086} (\bibinfo {year} {2013})}\BibitemShut {NoStop}%
\bibitem [{\citenamefont {Siskins}\ \emph {et~al.}(2020)\citenamefont
  {Siskins}, \citenamefont {Lee}, \citenamefont {Manas-Valero}, \citenamefont
  {Coronado}, \citenamefont {Blanter}, \citenamefont {van~der Zant},\ and\
  \citenamefont {Steeneken}}]{Siskins2020}%
  \BibitemOpen
  \bibfield  {author} {\bibinfo {author} {\bibfnamefont {M.}~\bibnamefont
  {Siskins}}, \bibinfo {author} {\bibfnamefont {M.}~\bibnamefont {Lee}},
  \bibinfo {author} {\bibfnamefont {S.}~\bibnamefont {Manas-Valero}}, \bibinfo
  {author} {\bibfnamefont {E.}~\bibnamefont {Coronado}}, \bibinfo {author}
  {\bibfnamefont {Y.~M.}\ \bibnamefont {Blanter}}, \bibinfo {author}
  {\bibfnamefont {H.~S.~J.}\ \bibnamefont {van~der Zant}},\ and\ \bibinfo
  {author} {\bibfnamefont {P.~G.}\ \bibnamefont {Steeneken}},\ }\bibfield
  {title} {\bibinfo {title} {Magnetic and {E}lectronic {P}hase {T}ransitions
  {P}robed by {N}anomechanical {R}esonators},\ }\href
  {https://doi.org/10.1038/s41467-020-16430-2} {\bibfield  {journal} {\bibinfo
  {journal} {Nat. Commun.}\ }\textbf {\bibinfo {volume} {11}},\ \bibinfo
  {pages} {2698} (\bibinfo {year} {2020})}\BibitemShut {NoStop}%
\bibitem [{\citenamefont {Ye}\ \emph {et~al.}(2018)\citenamefont {Ye},
  \citenamefont {Lee},\ and\ \citenamefont {Feng}}]{Ye2018Therm}%
  \BibitemOpen
  \bibfield  {author} {\bibinfo {author} {\bibfnamefont {F.}~\bibnamefont
  {Ye}}, \bibinfo {author} {\bibfnamefont {J.}~\bibnamefont {Lee}},\ and\
  \bibinfo {author} {\bibfnamefont {P.~X.}\ \bibnamefont {Feng}},\ }\bibfield
  {title} {\bibinfo {title} {Electrothermally {T}unable {G}raphene {R}esonators
  {O}perating at {V}ery {H}igh {T}emperature up to 1200 {K}},\ }\href
  {https://doi.org/10.1021/acs.nanolett.7b04685} {\bibfield  {journal}
  {\bibinfo  {journal} {Nano Lett.}\ }\textbf {\bibinfo {volume} {18}},\
  \bibinfo {pages} {1678} (\bibinfo {year} {2018})}\BibitemShut {NoStop}%
\bibitem [{\citenamefont {Baumeister}(2004)}]{BaumeisterOCT}%
  \BibitemOpen
  \bibfield  {author} {\bibinfo {author} {\bibfnamefont {P.~W.}\ \bibnamefont
  {Baumeister}},\ }\href {https://doi.org/10.1117/3.548071} {\emph {\bibinfo
  {title} {Optical Coating Technology}}},\ Vol.\ \bibinfo {volume} {PM137}\
  (\bibinfo  {publisher} {SPIE - The International Society for Optical
  Engineering},\ \bibinfo {address} {Bellingham, Washington USA},\ \bibinfo
  {year} {2004})\BibitemShut {NoStop}%
\bibitem [{\citenamefont {Islam}\ \emph {et~al.}(2018)\citenamefont {Islam},
  \citenamefont {van~den Akker},\ and\ \citenamefont {Feng}}]{Islam2018Ani}%
  \BibitemOpen
  \bibfield  {author} {\bibinfo {author} {\bibfnamefont {A.}~\bibnamefont
  {Islam}}, \bibinfo {author} {\bibfnamefont {A.}~\bibnamefont {van~den
  Akker}},\ and\ \bibinfo {author} {\bibfnamefont {P.~X.}\ \bibnamefont
  {Feng}},\ }\bibfield  {title} {\bibinfo {title} {Anisotropic {T}hermal
  {C}onductivity of {S}uspended {B}lack {P}hosphorus {P}robed by
  {O}pto-{T}hermomechanical {R}esonance {S}pectromicroscopy},\ }\href
  {https://doi.org/10.1021/acs.nanolett.8b03333} {\bibfield  {journal}
  {\bibinfo  {journal} {Nano Lett.}\ }\textbf {\bibinfo {volume} {18}},\
  \bibinfo {pages} {7683 } (\bibinfo {year} {2018})}\BibitemShut {NoStop}%
\bibitem [{\citenamefont {Byrnes}(2016)}]{ByrnesMOC}%
  \BibitemOpen
  \bibfield  {author} {\bibinfo {author} {\bibfnamefont {S.~J.}\ \bibnamefont
  {Byrnes}},\ }\bibfield  {title} {\bibinfo {title} {Multilayer {O}ptical
  {C}alculations},\ }\href {http://arXiv:1603.02720v5} {\bibfield  {journal}
  {\bibinfo  {journal} {arXiv:1603.02720v5}\ } (\bibinfo {year}
  {2016})}\BibitemShut {NoStop}%
\bibitem [{\citenamefont {Kurek}\ \emph {et~al.}(2017)\citenamefont {Kurek},
  \citenamefont {Carnoy}, \citenamefont {Larsen}, \citenamefont {Nielsen},
  \citenamefont {Hansen}, \citenamefont {Rades}, \citenamefont {Schmid},\ and\
  \citenamefont {Boisen}}]{Kurek2017Heat}%
  \BibitemOpen
  \bibfield  {author} {\bibinfo {author} {\bibfnamefont {M.}~\bibnamefont
  {Kurek}}, \bibinfo {author} {\bibfnamefont {M.}~\bibnamefont {Carnoy}},
  \bibinfo {author} {\bibfnamefont {P.~E.}\ \bibnamefont {Larsen}}, \bibinfo
  {author} {\bibfnamefont {L.~H.}\ \bibnamefont {Nielsen}}, \bibinfo {author}
  {\bibfnamefont {O.}~\bibnamefont {Hansen}}, \bibinfo {author} {\bibfnamefont
  {T.}~\bibnamefont {Rades}}, \bibinfo {author} {\bibfnamefont
  {S.}~\bibnamefont {Schmid}},\ and\ \bibinfo {author} {\bibfnamefont
  {A.}~\bibnamefont {Boisen}},\ }\bibfield  {title} {\bibinfo {title}
  {Nanomechanical {I}nfrared {S}pectroscopy with {V}ibrating {F}ilters for
  {P}harmaceutical {A}nalysis},\ }\href
  {https://doi.org/10.1002/anie.201700052} {\bibfield  {journal} {\bibinfo
  {journal} {Angew. Chem., Int. Ed.}\ }\textbf {\bibinfo {volume} {56}},\
  \bibinfo {pages} {3901} (\bibinfo {year} {2017})}\BibitemShut {NoStop}%
\bibitem [{\citenamefont {Krauthammer}(2001)}]{Krauthammer}%
  \BibitemOpen
  \bibfield  {author} {\bibinfo {author} {\bibfnamefont {E.}~\bibnamefont
  {Krauthammer}, \bibfnamefont {Theodor;~Ventsel}},\ }\href
  {https://doi.org/10.1201/9780203908723} {\emph {\bibinfo {title} {Thin
  {P}lates and {S}hells: {T}heory, {A}nalysis, and {A}pplications}}},\ \bibinfo
  {edition} {1st}\ ed.\ (\bibinfo  {publisher} {CRC Press},\ \bibinfo {address}
  {Boca Raton, USA},\ \bibinfo {year} {2001})\BibitemShut {NoStop}%
\bibitem [{\citenamefont {Wong}\ \emph {et~al.}(2010)\citenamefont {Wong},
  \citenamefont {Annamalai}, \citenamefont {Wang},\ and\ \citenamefont
  {Palaniapan}}]{wong2010characterization}%
  \BibitemOpen
  \bibfield  {author} {\bibinfo {author} {\bibfnamefont {C.~L.}\ \bibnamefont
  {Wong}}, \bibinfo {author} {\bibfnamefont {M.}~\bibnamefont {Annamalai}},
  \bibinfo {author} {\bibfnamefont {Z.~Q.}\ \bibnamefont {Wang}},\ and\
  \bibinfo {author} {\bibfnamefont {M.}~\bibnamefont {Palaniapan}},\ }\bibfield
   {title} {\bibinfo {title} {Characterization of {N}anomechanical {G}raphene
  {D}rum {S}tructures},\ }\href
  {https://doi.org/10.1088/0960-1317/20/11/115029} {\bibfield  {journal}
  {\bibinfo  {journal} {J. Micromechan. Microeng.}\ }\textbf {\bibinfo {volume}
  {20}},\ \bibinfo {pages} {115029} (\bibinfo {year} {2010})}\BibitemShut
  {NoStop}%
\bibitem [{\citenamefont {Weber}\ \emph {et~al.}(2016)\citenamefont {Weber},
  \citenamefont {Guttinger}, \citenamefont {Noury}, \citenamefont
  {Vergara-Cruz},\ and\ \citenamefont {Bachtold}}]{weber2016force}%
  \BibitemOpen
  \bibfield  {author} {\bibinfo {author} {\bibfnamefont {P.}~\bibnamefont
  {Weber}}, \bibinfo {author} {\bibfnamefont {J.}~\bibnamefont {Guttinger}},
  \bibinfo {author} {\bibfnamefont {A.}~\bibnamefont {Noury}}, \bibinfo
  {author} {\bibfnamefont {J.}~\bibnamefont {Vergara-Cruz}},\ and\ \bibinfo
  {author} {\bibfnamefont {A.}~\bibnamefont {Bachtold}},\ }\bibfield  {title}
  {\bibinfo {title} {Force {S}ensitivity of {M}ultilayer {G}raphene
  {O}ptomechanical {D}evices},\ }\href {https://doi.org/10.1038/ncomms12496}
  {\bibfield  {journal} {\bibinfo  {journal} {Nat. Commun.}\ }\textbf {\bibinfo
  {volume} {7}},\ \bibinfo {pages} {12496} (\bibinfo {year}
  {2016})}\BibitemShut {NoStop}%
\bibitem [{\citenamefont {Sazonova}(2006)}]{sazonova2006thesis}%
  \BibitemOpen
  \bibfield  {author} {\bibinfo {author} {\bibfnamefont {V.~A.}\ \bibnamefont
  {Sazonova}},\ }\emph {\bibinfo {title} {A {T}unable {C}arbon {N}anotube
  {R}esonator}},\ \href {https://hdl.handle.net/1813/3205} {\bibinfo {type}
  {Dissertation}},\ \bibinfo  {school} {Cornell University} (\bibinfo {year}
  {2006})\BibitemShut {NoStop}%
\bibitem [{\citenamefont {Wu}\ and\ \citenamefont {Zhong}(2011)}]{wu2011swcnt}%
  \BibitemOpen
  \bibfield  {author} {\bibinfo {author} {\bibfnamefont {C.~C.}\ \bibnamefont
  {Wu}}\ and\ \bibinfo {author} {\bibfnamefont {Z.}~\bibnamefont {Zhong}},\
  }\bibfield  {title} {\bibinfo {title} {Capacitive {S}pring {S}oftening in
  {S}ingle-{W}alled {C}arbon {N}anotube {N}anoelectromechanical {R}esonators},\
  }\href {https://doi.org/10.1021/nl1039549} {\bibfield  {journal} {\bibinfo
  {journal} {Nano Lett.}\ }\textbf {\bibinfo {volume} {11}},\ \bibinfo {pages}
  {1448} (\bibinfo {year} {2011})}\BibitemShut {NoStop}%
\bibitem [{\citenamefont {Pande}\ \emph {et~al.}(2020)\citenamefont {Pande},
  \citenamefont {Siao}, \citenamefont {Chen}, \citenamefont {Lee},
  \citenamefont {Sankar}, \citenamefont {Chang}, \citenamefont {Chen},
  \citenamefont {Chang}, \citenamefont {Chou},\ and\ \citenamefont
  {Lin}}]{pande2020ultralow}%
  \BibitemOpen
  \bibfield  {author} {\bibinfo {author} {\bibfnamefont {G.}~\bibnamefont
  {Pande}}, \bibinfo {author} {\bibfnamefont {J.-Y.}\ \bibnamefont {Siao}},
  \bibinfo {author} {\bibfnamefont {W.-L.}\ \bibnamefont {Chen}}, \bibinfo
  {author} {\bibfnamefont {C.-J.}\ \bibnamefont {Lee}}, \bibinfo {author}
  {\bibfnamefont {R.}~\bibnamefont {Sankar}}, \bibinfo {author} {\bibfnamefont
  {Y.-M.}\ \bibnamefont {Chang}}, \bibinfo {author} {\bibfnamefont {C.-D.}\
  \bibnamefont {Chen}}, \bibinfo {author} {\bibfnamefont {W.-H.}\ \bibnamefont
  {Chang}}, \bibinfo {author} {\bibfnamefont {F.-C.}\ \bibnamefont {Chou}},\
  and\ \bibinfo {author} {\bibfnamefont {M.-T.}\ \bibnamefont {Lin}},\
  }\bibfield  {title} {\bibinfo {title} {Ultralow {S}chottky {B}arriers in
  {H-BN} {E}ncapsulated {M}onolayer {WS}e$_{2}$ {T}unnel {F}ield-{E}ffect
  {T}ransistors},\ }\href {https://doi.org/10.1021/acsami.0c01025} {\bibfield
  {journal} {\bibinfo  {journal} {ACS Appl. Mater. Interfaces}\ }\textbf
  {\bibinfo {volume} {12}},\ \bibinfo {pages} {18667} (\bibinfo {year}
  {2020})}\BibitemShut {NoStop}%
\bibitem [{\citenamefont {Hill}\ \emph {et~al.}(2018)\citenamefont {Hill},
  \citenamefont {Rigosi}, \citenamefont {Krylyuk}, \citenamefont {Tian},
  \citenamefont {Nguyen}, \citenamefont {Davydov}, \citenamefont {Newell},\
  and\ \citenamefont {Walker}}]{hill2018comprehensive}%
  \BibitemOpen
  \bibfield  {author} {\bibinfo {author} {\bibfnamefont {H.~M.}\ \bibnamefont
  {Hill}}, \bibinfo {author} {\bibfnamefont {A.~F.}\ \bibnamefont {Rigosi}},
  \bibinfo {author} {\bibfnamefont {S.}~\bibnamefont {Krylyuk}}, \bibinfo
  {author} {\bibfnamefont {J.}~\bibnamefont {Tian}}, \bibinfo {author}
  {\bibfnamefont {N.~V.}\ \bibnamefont {Nguyen}}, \bibinfo {author}
  {\bibfnamefont {A.~V.}\ \bibnamefont {Davydov}}, \bibinfo {author}
  {\bibfnamefont {D.~B.}\ \bibnamefont {Newell}},\ and\ \bibinfo {author}
  {\bibfnamefont {A.~R.~H.}\ \bibnamefont {Walker}},\ }\bibfield  {title}
  {\bibinfo {title} {Comprehensive {O}ptical {C}haracterization of {A}tomically
  {T}hin {N}b{S}e$_{2}$},\ }\href {https://doi.org/10.1103/PhysRevB.98.165109}
  {\bibfield  {journal} {\bibinfo  {journal} {Phys. Rev. B.}\ }\textbf
  {\bibinfo {volume} {98}},\ \bibinfo {pages} {165109} (\bibinfo {year}
  {2018})}\BibitemShut {NoStop}%
\bibitem [{\citenamefont {Weber}\ \emph {et~al.}(2010)\citenamefont {Weber},
  \citenamefont {Calado},\ and\ \citenamefont {van~de Sanden}}]{Weber2010Grnk}%
  \BibitemOpen
  \bibfield  {author} {\bibinfo {author} {\bibfnamefont {J.~W.}\ \bibnamefont
  {Weber}}, \bibinfo {author} {\bibfnamefont {V.~E.}\ \bibnamefont {Calado}},\
  and\ \bibinfo {author} {\bibfnamefont {M.~C.~M.}\ \bibnamefont {van~de
  Sanden}},\ }\bibfield  {title} {\bibinfo {title} {Optical {C}onstants of
  {G}raphene {M}easured by {S}pectroscopic {E}llipsometry},\ }\href
  {https://doi.org/10.1063/1.3475393} {\bibfield  {journal} {\bibinfo
  {journal} {Appl. Phys. Lett.}\ }\textbf {\bibinfo {volume} {97}},\ \bibinfo
  {pages} {1} (\bibinfo {year} {2010})}\BibitemShut {NoStop}%
\bibitem [{\citenamefont {Hsu}\ \emph {et~al.}(2019)\citenamefont {Hsu},
  \citenamefont {Frisenda}, \citenamefont {Schmidt}, \citenamefont {Arora},
  \citenamefont {Vasconcellos}, \citenamefont {Bratschitsch}, \citenamefont
  {Zant},\ and\ \citenamefont
  {Castellanos‐Gomez}}]{Hsu2019Thickness-Dependent}%
  \BibitemOpen
  \bibfield  {author} {\bibinfo {author} {\bibfnamefont {C.}~\bibnamefont
  {Hsu}}, \bibinfo {author} {\bibfnamefont {R.}~\bibnamefont {Frisenda}},
  \bibinfo {author} {\bibfnamefont {R.}~\bibnamefont {Schmidt}}, \bibinfo
  {author} {\bibfnamefont {A.}~\bibnamefont {Arora}}, \bibinfo {author}
  {\bibfnamefont {S.~M.}\ \bibnamefont {Vasconcellos}}, \bibinfo {author}
  {\bibfnamefont {R.}~\bibnamefont {Bratschitsch}}, \bibinfo {author}
  {\bibfnamefont {H.~S.~J.}\ \bibnamefont {Zant}},\ and\ \bibinfo {author}
  {\bibfnamefont {A.}~\bibnamefont {Castellanos‐Gomez}},\ }\bibfield  {title}
  {\bibinfo {title} {Thickness‐{D}ependent {R}efractive {I}ndex of 1{L},
  2{L}, and 3{L} {M}o{S}$_{2}$, {M}o{S}e$_{2}$, {WS}$_{2}$, and {WS}e$_{2}$},\
  }\bibfield  {journal} {\bibinfo  {journal} {Adv. Opt. Mater.}\ }\textbf
  {\bibinfo {volume} {7}},\ \href {https://doi.org/10.1002/adom.201900239}
  {10.1002/adom.201900239} (\bibinfo {year} {2019})\BibitemShut {NoStop}%
\bibitem [{\citenamefont {Wang}\ and\ \citenamefont
  {Lan}(2016)}]{Wang2016AdvancesBP}%
  \BibitemOpen
  \bibfield  {author} {\bibinfo {author} {\bibfnamefont {X.}~\bibnamefont
  {Wang}}\ and\ \bibinfo {author} {\bibfnamefont {S.}~\bibnamefont {Lan}},\
  }\bibfield  {title} {\bibinfo {title} {Optical {P}roperties of {B}lack
  {P}hosphorus},\ }\bibfield  {journal} {\bibinfo  {journal} {Adv. Opt.
  Photonics}\ }\textbf {\bibinfo {volume} {8}},\ \href
  {https://doi.org/10.1364/aop.8.000618} {10.1364/aop.8.000618} (\bibinfo
  {year} {2016})\BibitemShut {NoStop}%
\bibitem [{\citenamefont {Palik}\ and\ \citenamefont
  {Prucha}(1997)}]{1997EPalik1}%
  \BibitemOpen
  \bibfield  {author} {\bibinfo {author} {\bibfnamefont {E.~D.}\ \bibnamefont
  {Palik}}\ and\ \bibinfo {author} {\bibfnamefont {E.~J.}\ \bibnamefont
  {Prucha}},\ }\href
  {https://doi.org/https://doi.org/10.1016/B978-012544415-6.50000-5} {\emph
  {\bibinfo {title} {Handbook of {O}ptical {C}onstants of {S}olids}}}\
  (\bibinfo  {publisher} {Academic Press},\ \bibinfo {address} {Burlington},\
  \bibinfo {year} {1997})\BibitemShut {NoStop}%
\bibitem [{\citenamefont {Zhou}\ \emph {et~al.}(2020)\citenamefont {Zhou},
  \citenamefont {Moldovan}, \citenamefont {Stan}, \citenamefont {Cai},
  \citenamefont {Czaplewski},\ and\ \citenamefont
  {Lopez}}]{Zhou2020StrainFree}%
  \BibitemOpen
  \bibfield  {author} {\bibinfo {author} {\bibfnamefont {J.}~\bibnamefont
  {Zhou}}, \bibinfo {author} {\bibfnamefont {N.}~\bibnamefont {Moldovan}},
  \bibinfo {author} {\bibfnamefont {L.}~\bibnamefont {Stan}}, \bibinfo {author}
  {\bibfnamefont {H.}~\bibnamefont {Cai}}, \bibinfo {author} {\bibfnamefont
  {D.~A.}\ \bibnamefont {Czaplewski}},\ and\ \bibinfo {author} {\bibfnamefont
  {D.}~\bibnamefont {Lopez}},\ }\bibfield  {title} {\bibinfo {title}
  {Approaching the {S}train-{F}ree {L}imit in {U}ltrathin {N}anomechanical
  {R}esonators},\ }\href {https://doi.org/10.1021/acs.nanolett.0c01027}
  {\bibfield  {journal} {\bibinfo  {journal} {Nano Lett.}\ }\textbf {\bibinfo
  {volume} {20}},\ \bibinfo {pages} {5693} (\bibinfo {year}
  {2020})}\BibitemShut {NoStop}%
\bibitem [{\citenamefont {Gil-Santos}\ \emph {et~al.}(2013)\citenamefont
  {Gil-Santos}, \citenamefont {Ramos}, \citenamefont {Pini}, \citenamefont
  {Llorens}, \citenamefont {Fernández-Regúlez}, \citenamefont {Calleja},
  \citenamefont {Tamayo},\ and\ \citenamefont {San~Paulo}}]{Santos2013SiNW}%
  \BibitemOpen
  \bibfield  {author} {\bibinfo {author} {\bibfnamefont {E.}~\bibnamefont
  {Gil-Santos}}, \bibinfo {author} {\bibfnamefont {D.}~\bibnamefont {Ramos}},
  \bibinfo {author} {\bibfnamefont {V.}~\bibnamefont {Pini}}, \bibinfo {author}
  {\bibfnamefont {J.}~\bibnamefont {Llorens}}, \bibinfo {author} {\bibfnamefont
  {M.}~\bibnamefont {Fernández-Regúlez}}, \bibinfo {author} {\bibfnamefont
  {M.}~\bibnamefont {Calleja}}, \bibinfo {author} {\bibfnamefont
  {J.}~\bibnamefont {Tamayo}},\ and\ \bibinfo {author} {\bibfnamefont
  {A.}~\bibnamefont {San~Paulo}},\ }\bibfield  {title} {\bibinfo {title}
  {Optical {B}ack-{A}ction in {S}ilicon {N}anowire {R}esonators: {B}olometric
  versus {R}adiation {P}ressure {E}ffects},\ }\bibfield  {journal} {\bibinfo
  {journal} {New J. Phys.}\ }\textbf {\bibinfo {volume} {15}},\ \href
  {https://doi.org/10.1088/1367-2630/15/3/035001}
  {10.1088/1367-2630/15/3/035001} (\bibinfo {year} {2013})\BibitemShut
  {NoStop}%
\bibitem [{\citenamefont {Wang}\ and\ \citenamefont
  {Feng}(2016)}]{wang2016interferometric}%
  \BibitemOpen
  \bibfield  {author} {\bibinfo {author} {\bibfnamefont {Z.}~\bibnamefont
  {Wang}}\ and\ \bibinfo {author} {\bibfnamefont {P.~X.}\ \bibnamefont
  {Feng}},\ }\bibfield  {title} {\bibinfo {title} {Interferometric {M}otion
  {D}etection in {A}tomic {L}ayer {2D} {N}anostructures: {V}isualizing {S}ignal
  {T}ransduction {E}fficiency and {O}ptimization {P}athways},\ }\href
  {https://doi.org/10.1038/srep28923} {\bibfield  {journal} {\bibinfo
  {journal} {Sci. Rep.}\ }\textbf {\bibinfo {volume} {6}},\ \bibinfo {pages}
  {28923} (\bibinfo {year} {2016})}\BibitemShut {NoStop}%
\bibitem [{\citenamefont {Li}\ and\ \citenamefont {Chen}(2017)}]{Li2017FIB}%
  \BibitemOpen
  \bibfield  {author} {\bibinfo {author} {\bibfnamefont {Z.}~\bibnamefont
  {Li}}\ and\ \bibinfo {author} {\bibfnamefont {F.}~\bibnamefont {Chen}},\
  }\bibfield  {title} {\bibinfo {title} {Ion {B}eam {M}odification of
  {T}wo-{D}imensional {M}aterials: {C}haracterization, {P}roperties, and
  {A}pplications},\ }\bibfield  {journal} {\bibinfo  {journal} {Appl. Phys.
  Rev.}\ }\textbf {\bibinfo {volume} {4}},\ \href
  {https://doi.org/10.1063/1.4977087} {10.1063/1.4977087} (\bibinfo {year}
  {2017})\BibitemShut {NoStop}%
\bibitem [{\citenamefont {Tomori}\ \emph {et~al.}(2019)\citenamefont {Tomori},
  \citenamefont {Hoshi}, \citenamefont {Inoue},\ and\ \citenamefont
  {Kanda}}]{Tomori2019FIB}%
  \BibitemOpen
  \bibfield  {author} {\bibinfo {author} {\bibfnamefont {H.}~\bibnamefont
  {Tomori}}, \bibinfo {author} {\bibfnamefont {N.}~\bibnamefont {Hoshi}},
  \bibinfo {author} {\bibfnamefont {D.}~\bibnamefont {Inoue}},\ and\ \bibinfo
  {author} {\bibfnamefont {A.}~\bibnamefont {Kanda}},\ }\bibfield  {title}
  {\bibinfo {title} {Influence of {F}ocused-{I}on-{B}eam {M}icrofabrication on
  {S}uperconducting {T}ransition in {E}xfoliated {T}hin {F}ilms of {L}ayered
  {S}uperconductor {N}b{S}e$_{2}$},\ }\bibfield  {journal} {\bibinfo  {journal}
  {Journal of Physics: Conference Series}\ }\textbf {\bibinfo {volume}
  {1293}},\ \href {https://doi.org/10.1088/1742-6596/1293/1/012006}
  {10.1088/1742-6596/1293/1/012006} (\bibinfo {year} {2019})\BibitemShut
  {NoStop}%
\end{thebibliography}

\providecommand{\noopsort}[1]{}\providecommand{\singleletter}[1]{#1}%

\end{document}